\documentclass[aps,10pt,prb,twocolumn,amssymb,amsmath,amsfonts,showpacs,floatfix,citeautoscript,preprintnumbers,footinbib,a4paper]{revtex4-1}
\usepackage{times}
\usepackage[colorlinks,bookmarks=false,citecolor=blue,linkcolor=red,urlcolor=blue]{hyperref}
\usepackage{graphicx,epstopdf}
\usepackage[latin1]{inputenc}

\newcommand{\spin}[1]{\mathbf{S}_{#1}}
\newcommand{\opspin}[1]{\hat{\mathbf{S}}_{#1}}
\newcommand{\op}[1]{\hat{#1}}

\newcommand{\moy}[1]{\langle{#1}\rangle}

\newcommand{\Ham}{\mathcal{H}}

\newcommand{\Heff}[1]{\mathbf{H}^{\text{eff}}_\mathbf{#1}}
\newcommand{\Jeff}[1]{J^{\text{eff}}_{#1}}
\newcommand{\doping}{z}

\begin{document}

\title{Magnetic responses of randomly depleted spin ladders}

\author{Arthur Lavar\'elo}\email{arthur.lavarelo@u-psud.fr}
\author{Guillaume Roux}\email{guillaume.roux@u-psud.fr}
\affiliation{Laboratoire de Physique Th\'eorique et Mod\`eles statistiques, Universit\'e Paris-Sud, CNRS, UMR8626, 91405 Orsay, France.}

\author{Nicolas Laflorencie}\email{nicolas.laflorencie@irsamc.ups-tlse.fr}
\affiliation{Laboratoire de Physique Th\'eorique, Universit\'e de Toulouse, UPS, (IRSAMC), Toulouse, France}

\date{\today}

\begin{abstract}
  The magnetic responses of a spin-$1/2$ ladder doped with
  non-magnetic impurities are studied combining both analytical and
  numerical methods.  The regime where frustration induces
  incommensurability is taken into account.  Several improvements are
  made on the results of the seminal work by Sigrist and Furusaki
  [Sigrist and A.~Furusaki, J. Phys. Soc. Jpn., \textbf{65}, 2385
  (1996)], and deviations from the Brillouin magnetization curve due
  to interactions are also analyzed. We first discuss the magnetic
  profile around a single impurity and the effective interactions
  between impurities within the bond-operator mean-field theory. The
  results are compared to density-matrix renormalization group
  calculations. In particular, these quantities are shown to be
  sensitive to the transition to the incommensurate regime. We then
  focus on the behavior of the zero-field susceptibility through an
  effective Curie constant. At zero-temperature, we give
  doping-dependent corrections to the results of Sigrist and Furusaki
  on general bipartite lattices, and compute exactly the distribution
  of ladder clusters due to chain breaking effects. Solving the
  effective model with exact diagonalization and quantum Monte-Carlo
  gives the temperature dependence of the Curie constant. Its
  high-temperature limit is understood within a random dimer model,
  while the low-temperature tail is compatible with a real-space
  renormalization group scenario.  Interestingly, solving the full
  microscopic model does not show a plateau corresponding to the
  saturation of the impurities \emph{in isotropic ladders}.  The
  second magnetic response which is analyzed is the magnetic curve.
  Below fields of the order of the spin gap, the magnetization process
  is controlled by the physics of interacting impurity spins.  The
  random dimer model is shown to capture the bulk of the curve,
  accounting for the deviation from a Brillouin behavior due to
  interactions.  The effective model calculations agree rather well
  with density-matrix renormalization group calculations at zero
  temperature, and with quantum Monte-Carlo at finite temperature. In
  all, the effect of incommensurability does not display a strong
  qualitative effect on both the magnetic susceptibility and the
  magnetic curve. Consequences for experiments on the BiCu$_2$PO$_6$
  compound and other spin-gapped materials are briefly mentioned.
\end{abstract}


\maketitle

\section{Introduction}

The presence of disorder and impurities in strongly correlated systems
offers a good opportunity to better understand the role played by
quantum fluctuations in such materials~\cite{Alloul2009}. Either
intrinsically present or explicitly added by doping, impurities in
condensed matter systems can rarely be ignored, in particular when
they induce new physics as compared to the disorder-free
situation. Prominent examples are the Kondo effect~\cite{Hewson1997},
Anderson localization~\cite{Evers2008}, dirty bosons physics in
disordered superconductors~\cite{Fisher1989}, impurities in magnetic
semiconductors~\cite{Sato10}, spin glasses~\cite{Binder1986}...

In the context of antiferromagnetic (AF) Mott insulators, parent
compounds of high temperature superconducting cuprates for instance,
spin ladder materials~\cite{Dagotto1996} have been shown to display
very interesting features, in particular when the number of legs is an
even number. For example, a finite energy gap $\Delta_s$ appears in
the excitation spectrum of two-leg AF spin-1/2
ladders~\cite{Johnston87,Barnes1993,Gopalan1994}, as seen for instance
in SrCu$_2$O$_3$~\cite{Azuma94}. Furthermore, defects in gapped
ladders induce very interesting
effects~\cite{Azuma1997,Azuma1998,Fujiwara1998,Ohsugi1999,Koteswararao2010,Casola2010},
in particular the apparition of effective gapless modes below the bare
spin gap $\Delta_s$. Having in mind that the ground-state of a two-leg
ladder displays short-range resonating valence bond like
physics~\cite{White96}, a non-magnetic dopant is expected to break
such a short-distance singlet, inducing a quasi-free
spin-$\frac{1}{2}$, strongly localized in the vicinity of the
impurity. Interesting questions arise then when a finite concentration
of impurities is introduced in a spin ladder, as studied in a large
number of theoretical works~\cite{Fukuyama1996,Sigrist1996,
  Nagaosa1996,
  Motome1996,Iino1996,Imada1997,Mikeska1997,Sandvik1997,Miyazaki1997,Martins1997,
  Laukamp1998,Greven1998,Arlego2004,Anfuso2006, Trinh2012}. Similar
physics is also at play in other spin-gapped materials: spin-$1$
(Haldane) chains such as Y$_2$BaNiO$_5$~\cite{Tedoldi99,Das04}, or
PbNi1$_2$V$_2$O$_8$~\cite{Imai04}, spin-Peierls chains such as
CuGeO$_3$~\cite{Hase93,Fukuyama96,Manabe98,Grenier98,Augier1999,
  Shelton1998, Normand2002, Laflorencie2004, Laflorencie2005} for
instance. Indeed, a universal behavior can be observed for the
impurity-induced three-dimensional ordering mechanism in such weakly
coupled chains or ladder materials~\cite{Bobroff09}.

Nevertheless, several aspects of impurity effects in ladder -- and
more generally spin gapped -- materials remain to be explored in order
to better understand and interpret experimental results. Regarding the
effective pair-wise interaction between released moments, it is
believed to remain non-frustrated even when the underlying spin
systems is frustrated~\cite{Dobry99,Laflorencie03}, but it is not
clear to which extend such a result is robust when strong frustration
leads to incommensurability~\cite{Andrade2012}, as expected for
instance in the ladder material
BiCu$_2$PO$_6$~\cite{Mentre2009,Tsirlin2010, Alexander2010,
  Koteswararao2010,Lavarelo2011,Sugimoto2013,Plumb2013,Choi2013}. A
natural question then arises regarding which effective model is able
to quantitatively describe the low energy physics of randomly doped
ladders. Indeed, it was believed since the seminal work of Sigrist and
Furusaki~\cite{Sigrist1996} that a simple model of random (in sign and
magnitude, reflecting the random locations of impurities in a ladder)
nearest-neighbors couplings between effective spin-1/2 (describing
impurity degrees of freedom) was able to correctly capture the low
temperature physics of depleted ladders. This so-called random F-AF
chain model~\cite{Furusaki94,Furusaki1995, Westerberg1997,
  Frischmuth1997, Frischmuth1999, Hikihara1999,Yusuf2003,Hoyos04}
displays some universal behavior for various quantities such as
uniform and staggered susceptibilities or the specific heat in the low
temperature regime, with an interesting large spin phase occurring at
very low temperature. However, in the context of depleted ladders,
universality for such thermodynamic quantities has been first
questioned using quantum Monte Carlo (QMC) simulations by Miyazaki and
co-workers~\cite{Miyazaki1997} where no clear signature of universal
low temperature scalings were found, in agreement with a more recent
QMC study~\cite{Trinh2012}.

Despite the large number of works devoted to such systems in the
absence of external magnetic field, much less is known regarding
finite field effects. Indeed, as recently reviewed by Giamarchi and
co-workers~\cite{Giamarchi2008}, applying a finite external field on
gapped AF systems leads to the analog of a Bose-Einstein condensation
(BEC) of magnetic excitations~\cite{Affleck1991,Giamarchi99}
(hard-core bosons triplets) when the field is sufficiently strong to
close the spin gap $\Delta_s$. Note that a true BEC is only expected
for dimension $d> 2$, which occurs at low enough temperature below
some energy scale controlled by three dimensional
couplings. Nevertheless for low-$d$ a quasi-BEC is expected, as
observed in ultra-cold atom physics~\cite{Petrov00,Dettmer01}. In
solid state physics, triplet BEC has been observed in several quantum
magnetic compounds, such as coupled dimers
TlCuCl$_3$~\cite{Nikuni00,Ruegg03}, frustrated bilayers
BaCuSi$_2$O$_6$~\cite{Jaime04}, coupled Haldane chains in
DTN~\cite{Zapf06}, and also spin ladder materials like
(C$_5$H$_{12}$N)$_2$CuBr$_4$~\cite{Klanjsek08,Ruegg08,Thielemann09}.

However, when disorder is present in such spin gapped systems, a new
phenomenology is expected with the interesting possibility to address
Bose-Glass (BG) physics, as recently found in Br-doped
IPaCuCl$_3$~\cite{Hong10} or DTN~\cite{Yu2012} (see also
Ref.~\onlinecite{Zheludev2013} for a very recent review). While
several issues remain unsolved regarding BG physics, {\it{e.g.}} for
the excitation spectrum~\cite{Roux13,Vojta13,Alvarez13}, the case
where disorder comes from ligand substitution seems easier to
understand from a microscopic point of view. Indeed, such doping will
essentially generate disorder in the AF couplings without inducing
local moments. On the other hand, doping on the magnetic sites is
expected to be more complicated as gapless states will populate the
clean gap. Therefore the magnetic response will display non-trivial
Brillouin-like behaviors in most of the experimentally relevant
situations. Such cases have been studied theoretically by a few
authors~\cite{Mikeska2004, Roscilde2006, Yu2010a, Yu2010}, showing a
rich physics and various scenarios that demand further analysis.

In this work, we focus on the two-leg ladder model to provide a
systematic analysis of the physics of interacting impurities, building
on both analytical and numerical arguments.  In particular, we are
interesting in the following issues: (i) the effective interaction
between impurities for commensurate and incommensurate backgrounds;
(ii) the low energy emergence of large spins due to random signs in
effective couplings in a realistic context including finite size
effects due to chain breaking; (iii) the temperature scaling of the
Curie constant of the uniform susceptibility, as obtained from both
effective and realistic doped ladder models; (iv) the deviations of
the magnetic curve from the Brillouin response as a probe of the
magnitude of interactions.  The ladder model used throughout this
study is the one studied in Refs.~\onlinecite{Vekua2006, Lavarelo2011}
\begin{align}
\nonumber
\mathcal{H}=\sum_{i=1}^L\quad  &J_{1}\left[\mathbf{S}_{i,1}\cdot\mathbf{S}_{i+1,1} + \mathbf{S}_{i,2}\cdot\mathbf{S}_{i+1,2}\right]\\
\nonumber
      + &J_{2} \left[\mathbf{S}_{i,1}\cdot\mathbf{S}_{i+2,1} + \mathbf{S}_{i,2}\cdot\mathbf{S}_{i+2,2}\right]\\
\label{eq:microH}
      + &J_{\perp} \,\mathbf{S}_{i,1}\cdot\mathbf{S}_{i,2} \;,
\end{align}
where $\mathbf{S}_{i,j}$ is the spin-$1/2$ operator acting at site $i$
of leg $j$ and the $J$s are the magnitude of the various couplings
which are here taken to be antiferromagnetic ($J>0$). In the rest of
the paper, the only parameter coming with the presence of impurities
is their concentration $\doping$. The doped microscopic model is
numerically solved with two state-of-the-art methods: the
density-matrix renormalization group (DMRG) technique~\cite{DMRG} and
the stochastic series expansion (SSE) quantum Monte-Carlo (QMC)
technique~\cite{QMC}.

The paper is organized as follow: In a first part, we discuss in
details the effective model describing two-body interactions between
impurities. The resulting effective model is then compared to the
solution of the microscopic model in the second part. The latter is
dedicated to the study of the magnetization curve at field below the
spin gap $\Delta_s$, i.e. in the region dominated by the impurity
spins response.  This region is itself divide in two regimes: (i) The
small field regime $H\ll T$, featuring a temperature-dependent Curie
constant $c(T)$, (ii) the intermediate field regime $T \lesssim H
\lesssim \Delta_s$ displaying again deviations from Brillouin through
an approximate power-law behavior. We do not investigate fields $H
\gtrsim \Delta_s$ as the physics involves triplet bosons in a
disordered medium which is exciting but beyond the scope of the
present manuscript.

\section{Effective interaction between impurities}

In this first section, the emphasis is put on the quantitative
analysis of the effective interaction between impurities from
arguments similar to RKKY theory. This provides an effective
Hamiltonian which couplings distribution is essential for
understanding the magnetic responses. Previous works along this
direction are found in Refs.~\onlinecite{Sigrist1996, Nagaosa1996}.

\subsection{Effective Hamiltonian}

We start with the derivation of the low-energy effective Hamiltonian
accounting for effective interactions between impurities.  For a
generic Heisenberg spin model with $N$ spins, the clean Hamiltonian
takes the general form
\begin{equation}
 \Ham_\text{clean}=\frac{1}{2}\sum_{\mathbf{r},\mathbf{r}'} J_{\mathbf{r}-\mathbf{r}'} \spin{\mathbf{r}}\cdot\spin{\mathbf{r}'}\;,
\end{equation}
where $J_{\mathbf{R}}$ are the microscopic couplings, which depend
only on the relative distance $\mathbf{R}= \mathbf{r}-\mathbf{r}'$.
We now consider that a few non-magnetic impurities occupy sites
$\textbf{I}$ of the lattice.  Then, the Hamiltonian reads
\begin{equation}
 \Ham=\frac{1}{2}\sum_{\mathbf{r}\neq\textbf{I}}\sum_{\mathbf{r}'\neq\textbf{I}} J_{\mathbf{r}-\mathbf{r}'} \spin{\mathbf{r}}\cdot\spin{\mathbf{r}'}\;,
\end{equation}
or $\Ham=\Ham_\text{clean}+\Ham_\text{imp}$, where
\begin{equation}
\label{eq:Himp}
 \Ham_\text{imp}=-\sum_{\mathbf{r}}\sum_{\mathbf{I}} J_{\mathbf{r}-\mathbf{I}} \spin{\mathbf{r}}\cdot\spin{\mathbf{I}}\;.
\end{equation}
Notice that effective spins operators are introduced at sites
$\textbf{I}$ where the impurities live, while these sites are actually
vacant. Hamiltonian \eqref{eq:Himp} takes the form $
\Ham_\text{imp}=-\sum_\mathbf{r} \Heff{r}\cdot\spin{\mathbf{r}}$, in
which $\Heff{r} = \sum_{\mathbf{I}} J_{\mathbf{r}-\mathbf{I}}
\spin{\mathbf{I}}$ is an effective magnetic field operator. Assuming
the perturbation $\Ham_\text{imp}$ can be treated using linear
response theory, we may write the Fourier transform of
$\spin{\mathbf{r}}$:
\begin{equation}
\spin{\mathbf{k}} \simeq \chi_\mathbf{k}\Heff{k}\;,
\end{equation}
with $\chi_\mathbf{k}$
is the static susceptibility at wave-vector $\mathbf{k}$, and
\begin{equation}
\Heff{k} = J_{\mathbf{k}}\sum_{\mathbf{I}}\spin{\mathbf{I}} e^{i\mathbf{k}\cdot\mathbf{I}}\;.
\label{eq:Heff_Fourier}
\end{equation}
One can thus write the perturbation $\Ham_\text{imp}$ as
\begin{equation}
\Ham_\text{imp}= \sum_{\mathbf{I},\mathbf{J}} \Jeff{\mathbf{I}-\mathbf{J}} \spin{\mathbf{I}}\cdot\spin{\mathbf{J}}\;,
\label{eq:Himp-realspace}
\end{equation}
in which
\begin{equation}
\Jeff{\mathbf{R}} = -\sum_{\mathbf{k}} |J_{\mathbf{k}}|^2 \chi_\mathbf{k} e^{-i\mathbf{k}\cdot\mathbf{R}}
\label{eq:Jeff-realspace}
\end{equation}
is the effective two-body interaction between impurities.

When the clean system possesses a spin gap associated to a spin
correlation length $\xi_\text{spin}$, the susceptibility
$\chi_\mathbf{R}$, and therefore the effective interaction
$\Jeff{\mathbf{R}}$, decreases exponentially with the distance
$\lVert\mathbf{R}\rVert$. For a sufficiently small impurity
concentration $\doping$ ($\doping\ll 1/\xi_\text{spin}$ in one
dimension), effective interactions remain much smaller than the spin
gap. At temperatures smaller than this gap, the clean part of the
doped system can be considered to be in the ground-state of
$\Ham_\text{clean}$ while the impurities dynamics is governed by
\eqref{eq:Himp-realspace}, in which one can take the zero-temperature
behavior for the susceptibility $\chi_\mathbf{k}$.

\subsection{Static susceptibility within the BOMF approximation}

The static susceptibility of the ground-state of \eqref{eq:microH} can
be computed using the bond-order mean-field (BOMF)
approximation~\cite{Gopalan1994, Lavarelo2011} (see
Appendix.~\ref{app:BOMF}). In the strong-coupling limit $J_{\perp}\gg
J_1$, the spin gap is in the $k_y=\pi$ sector and the magnon branch is
well separated from the two-magnons continuum.  On can thus neglect
the $k_y=0$ contribution and keep only the single magnon one.  The
details of the calculations are given in Appendix~\ref{app:BOMF} and
show that the susceptibility displays the same singularities as the
spin structure factor.  In the large $J_\perp$ regime, the result
reads
\begin{equation}
\chi_{k,\pi} \simeq \frac{1}{4J_{\perp} + 8J_1 \cos{k} + 8J_2\cos{2k}} \;.
\end{equation}

\subsection{Magnetization profile induced by a single impurity}

Before turning to the interaction between two impurities, it is first
instructive to consider the magnetization pattern induced by a single
impurity, which is also of interest for nuclear magnetic resonance
(NMR) experiments.  The impurity is located at site $\mathbf{I}_0$ and
the corresponding effective magnetic field defined by
\eqref{eq:Heff_Fourier} is simply given by $\Heff{r} =
J_{\mathbf{r}-\mathbf{I}_0}\spin{\mathbf{I}_0}$. The expectation value
of the spin operator $\spin{\mathbf{r}}$ is then given by linear
response theory which, in Fourier transform, reads
\begin{equation}
\moy{\spin{\mathbf{k}}}\simeq \chi_\mathbf{k} \moy{\Heff{k}}\;.
\end{equation}
This perturbative response is a priori valid far from the
impurity. The magnetization profile in sector $S^z_\text{tot} = 1/2$
is then
\begin{equation}
\moy{S^z_{\mathbf{I}_0+\mathbf{R}}} \simeq \frac{1}{2\sqrt{2L}} \sum_\mathbf{k} e^{-i\mathbf{k}\cdot\mathbf{R}} \chi_\mathbf{k} J_{\mathbf{k}}\;,
\label{eq:mean-S-BOMF}
\end{equation}
in which the general expression of the coupling of the frustrated
ladder Hamiltonian \eqref{eq:microH} is
\begin{equation} 
J_{\mathbf{k}} = \frac{1}{\sqrt{2L}}\left(2J_1\cos k_x+2J_2\cos 2k_x+J_\perp \cos k_y\right)\;.
\end{equation}
After computing the integral limit of the sum \eqref{eq:mean-S-BOMF}
over the Brillouin zone, one obtains two different situations,
depending on the behavior of the residues: In the commensurate regime
$\frac{J_2}{J_1}<\frac{J_1}{4J_\perp}$, the profile is given by
\begin{widetext}
\begin{equation}
\label{eq:Sz_commensurable}
\moy{S^z_{x,y}}\simeq\frac18(-1)^{x+y}\Bigg( \frac{e^{-x/\xi_\text{spin}^+}}{\sinh\left(1/\xi_\text{spin}^+\right) P'\left[-\cosh\left(1/\xi_\text{spin}^+\right)\right]}
+\frac{e^{-x/\xi_\text{spin}^-}}{\sinh\left(1/\xi_\text{spin}^-\right) P'\left[-\cosh\left(1/\xi_\text{spin}^-\right)\right]}  \Bigg) \;,
\end{equation}
where $\xi_\text{spin}^\pm$ are the spin correlation lengths defined
in Eqs.~\eqref{eq:xi_commensurable} and $P'(X)$ is the derivative of
the polynomial $P(X)$ defined in Eq.~\eqref{eq:polynome}. In the
incommensurate regime $\frac{J_2}{J_1}>\frac{J_1}{4J_\perp}$, one has
\begin{equation}
\label{eq:Sz_incommensurable}
\moy{S^z_{x,y}}\simeq\frac14(-1)^{y}e^{-x/\xi_\text{spin}}\Im\Bigg[ \frac{e^{iqx}}{\sin\left(q+i\xi_\text{spin}^{-1}\right)P'\left[\cos\left(q+i\xi_\text{spin}^{-1}\right)\right]}
\Bigg] \;,
\end{equation}
\end{widetext}
where $q$ and $\xi_\text{spin}$ are defined by Eqs.~\eqref{eq:q} and
\eqref{eq:xi_incommensurable}. Notice that there is no unknown
constant in these expressions.

The key point of this result is that the transition from commensurate
to incommensurate correlations induces a discontinuity in the features
of the magnetization profile, which will show up in the effective
interaction too. Indeed, at each side of the transition, both residues
diverge but their sum tends to zero. Notice that exactly at the
transition, the denominator factorizes, having a single pole of order
2 for which the residue is zero, corresponding to
$\moy{S^z_{x,y}}=0$. On the contrary, the amplitude of the
magnetization $\moy{S^z_{x,y}}$ goes to $+\infty$ at each side of the
transition. But this does not mean that the magnetization profile
diverges at short distance. Of course, $\moy{S^z_{x,y}}$ remains
always bound by $1/2$. However, the fact that the amplitude of the
asymptotic behavior diverges which makes the perturbative analysis of
the linear response fail.

\begin{figure}[t]
\centering
\includegraphics[width=0.8\columnwidth,clip]{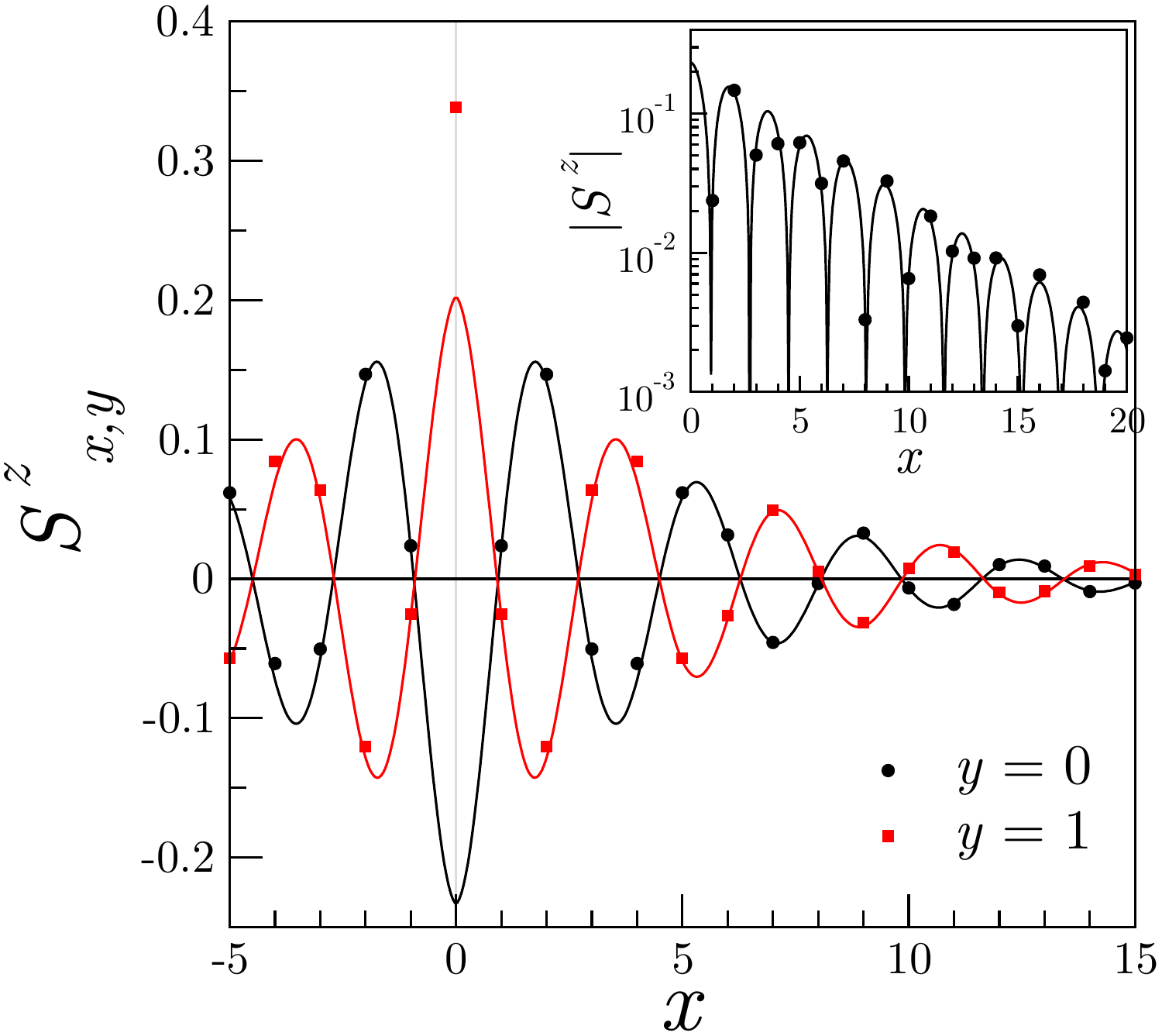}
\caption{(color online) Magnetization profile in sector
  $S^z_\text{tot}=1/2$ induced by a non-magnetic impurity at site
  $(0,0)$, for an isotropic ladder ($J_\perp=J_1$) with $J_2=J_1$. The
  incommensurate regime displays oscillations at wave-vector $q$.  Fit
  is done using \eqref{fit_incommensurable}.}
\label{fig:aimantation_impurete}
\end{figure}

On Fig.~\ref{fig:aimantation_impurete}, we compare the magnetization
profile in the sector $S^z_\text{tot}=1/2$ obtained by DMRG to the
mean-field predictions.  In practice, expressions
\eqref{eq:Sz_commensurable} and \eqref{eq:Sz_incommensurable} provide
good estimates of the behaviors, but it is preferable to fit the
magnetization profiles using the following ansatz:
\begin{equation}
\label{fit_commensurable}
\moy{S^z_{x,y}}=C(-1)^{x+y+1} e^{-x/\xi_\text{spin}}\;,
\end{equation}
in the commensurate regime, and 
\begin{equation}
\label{fit_incommensurable}
\moy{S^z_{x,y}}=C(-1)^{y+1} e^{-x/\xi_\text{spin}}\cos(qx+\phi)\;,
\end{equation}
in the incommensurate one. Remarkably, except near the onset of
incommensurability where the amplitude diverges, these expressions
remain correct at small distances, down to $x=1$. In NMR experiments,
an incommensurate $q$ would give a narrowing of the peak w.r.t. the
commensurate case with the same $\xi_{\text{spin}}$ since the
magnetization will display smaller values even close to the impurity.

\begin{figure}[t]
\centering
\includegraphics[width=\columnwidth,clip]{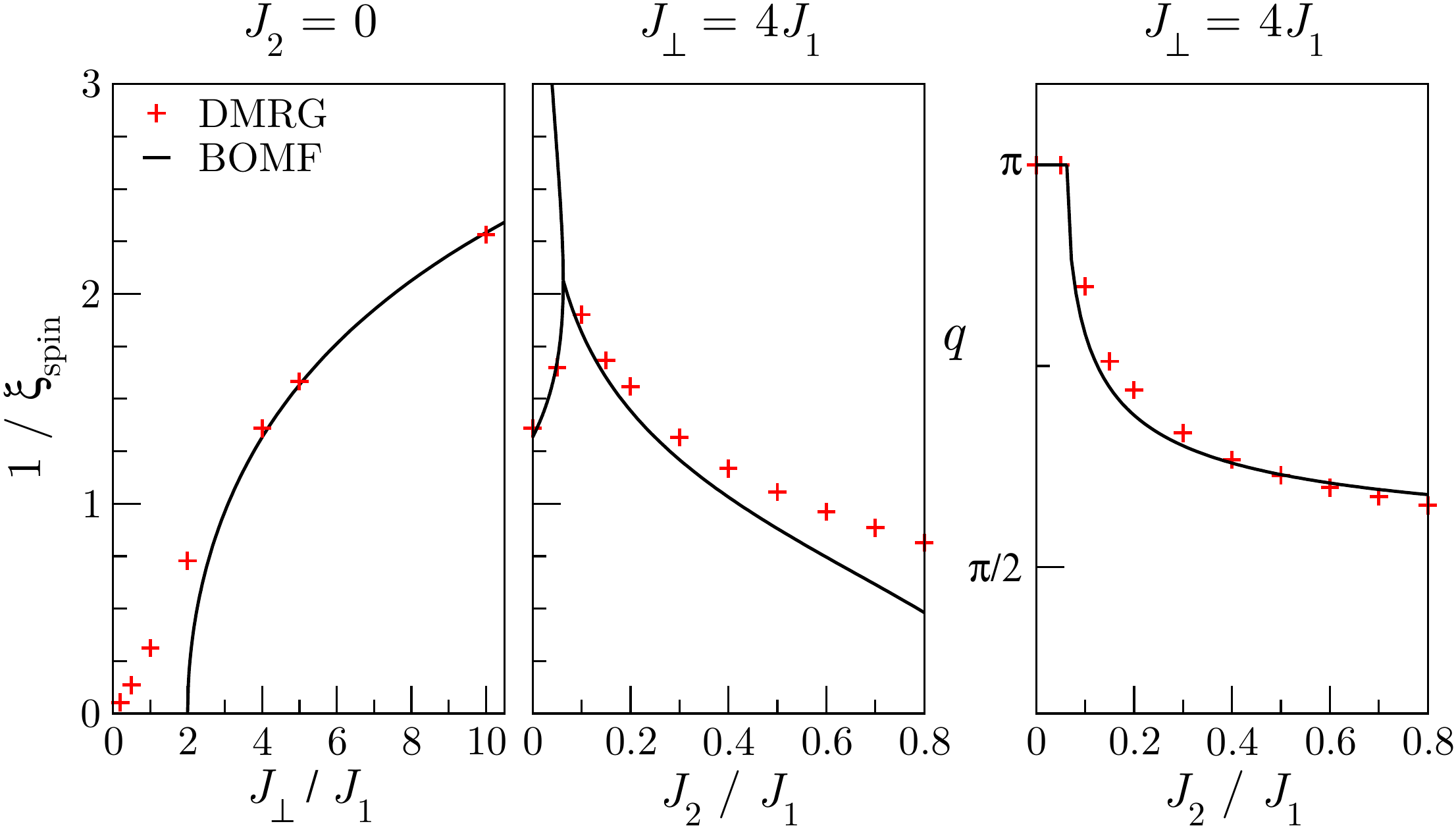}
\caption{(color online) Correlation length $\xi_\text{spin}$ and
  wave-vector $q$ extracted from the magnetization profile induced by
  a single impurity. DMRG results are compared to BOMF predictions in
  the large $J_\perp$ regime.}
\label{fig:xi_q}
\end{figure}

These profiles give a simple way to numerically access the fit
parameters and compare them to BOMF predictions. Indeed, the values of
$\xi_\text{spin}$ and $q$ extracted from the magnetization profiles
agree qualitatively well with the mean-field predictions, as shown on
Fig.~\ref{fig:xi_q}. In particular, we checked that the amplitude $C$
possesses a maximum close to the transition from commensurate to
incommensurate. Physically, these calculations provide an explicit
illustration of the fact that the impurity generates a spinon that is
confined close to it through an effective attractive potential acting
over a typical length-scale $\xi_\text{spin}$.

\begin{figure}[t]
\centering
\includegraphics[width=0.75\columnwidth,clip]{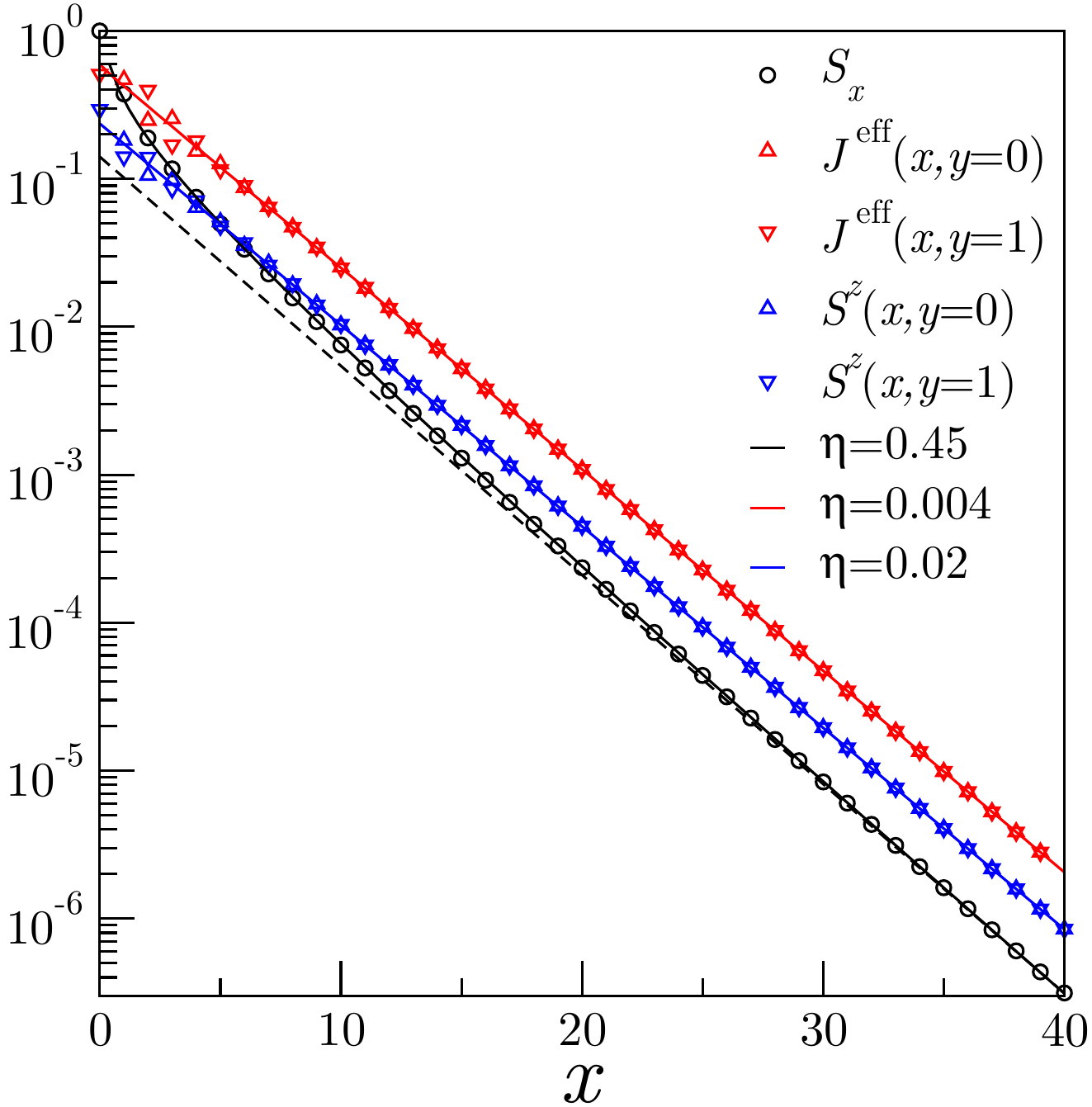}
\caption{(color online) Magnetization profile in sector
  $S^z_\text{tot}=1/2$ induced by a non-magnetic impurity at site
  $(0,0)$, for an isotropic ladder ($J_\perp=J_1$) for $J_2=0$ in the
  commensurate regime. The decay of the magnetization profile and
  effective interaction between two impurities are compared to the
  decay of spin correlations. These three quantities display the same
  length-scale $\xi_\text{spin}$ in the exponential. Yet, only
  correlations display a power-law correction (see
  App.~\ref{app:BOMF}).}
\label{fig:correction}
\end{figure}

Last, we stress that there is a qualitative difference between the
magnetization profile and the spin correlation function (see
App.~\ref{app:BOMF} for discussion of spin correlations in the model).
One does not expect a power-law correction in the decay of the
magnetization.  This is clearly visible on Fig~\ref{fig:correction}
where fitting the envelope using $e^{-x/\xi_\text{spin}}/x^\eta$ gives
an exponent $\eta\simeq 0$ while the exponent found for the fit of the
correlations is rather $\eta\simeq1/2$, as expected from the usual
arguments recalled in App.~\ref{app:BOMF}. These quantitative results
on the magnetization profiles and their sensitivity to the
commensurate-incommensurate transition share similarities with those
on the effective interaction between impurities which we now discuss.

\subsection{Effective interaction between impurities}

\subsubsection{Long-distance behavior}

Within the BOMF approximation, valid in the strong-coupling limit, the
effective interaction between impurities of
Eq.~\eqref{eq:Jeff-realspace} takes the following form in the
thermodynamical limit:
\begin{equation}
\Jeff{x,y}\simeq\frac{J_\perp}8 (-1)^{y+1} \frac{1}{2\pi} \int_0^{2\pi}\frac{Q^2(\cos k)}{P(\cos k)}e^{ikx}dk\;,
\end{equation}
where $Q$ is the polynom
\begin{equation}
 Q(X)=1+2\frac{J_2}{J_\perp}-2\frac{J_1}{J_\perp}X-4\frac{J_2}{J_\perp}X^2\;.
\end{equation}
As for the magnetization profile, one can evaluate the integral using
the residue theorem to obtain two cases:
\begin{widetext}
in the commensurate regime, one has
\begin{multline}
\label{eq:Jeff_commensurable}
\Jeff{x,y}\simeq\frac18(-1)^{x+y+1}\Bigg( \frac{Q^2\left[-\cosh\left(1/\xi_\text{spin}^+\right)\right]}{\sinh\left(1/\xi_\text{spin}^+\right) P'\left[-\cosh\left(1/\xi_\text{spin}^+\right)\right]}e^{-x/\xi_\text{spin}^+}
+\frac{Q^2\left[-\cosh\left(1/\xi_\text{spin}^-\right)\right]}{\sinh\left(1/\xi_\text{spin}^-\right) P'\left[-\cosh\left(1/\xi_\text{spin}^-\right)\right]} e^{-x/\xi_\text{spin}^-} \Bigg) \;,
\end{multline}
while in the incommensurate regime, one has
\begin{equation}
\label{eq:Jeff_incommensurable}
\Jeff{x,y}\simeq\frac14(-1)^{y+1}e^{-x/\xi_\text{spin}}\Im\Bigg[ \frac{Q^2\left[\cos\left(q+i\xi_\text{spin}^{-1}\right)\right]}{\sin\left(q+i\xi_\text{spin}^{-1}\right)P'\left[\cos\left(q+i\xi_\text{spin}^{-1}\right)\right]}e^{iqx}
\Bigg] \;.
\end{equation}
\end{widetext}
Similarly to the magnetization profile, the amplitude of $\Jeff{x,y}$
diverges close to the transition between the two regimes but is
strictly zero at the transition point. This divergence does not mean
that the effective interaction gets stronger but rather that the
applicability of the long-distance result is limited to large
distances. The effective interaction remains always bonded by the
maximum of $J_1$ and $J_2$ (see short distances behavior hereafter).

Further, no power-law corrections are expected in this quantity,
contrarily to what is commonly proposed~\cite{Sigrist1996,
  Roscilde2006}. Yet, this result is valid in the strong-coupling
limit and we observe that it remains correct down to the isotropic
ladder regime. In the weak-coupling limit, it is possible to have
power-law or logarithmic corrections but we have not studied this case
quantitatively.

\begin{figure}[t]
\centering
\includegraphics[width=\linewidth,clip]{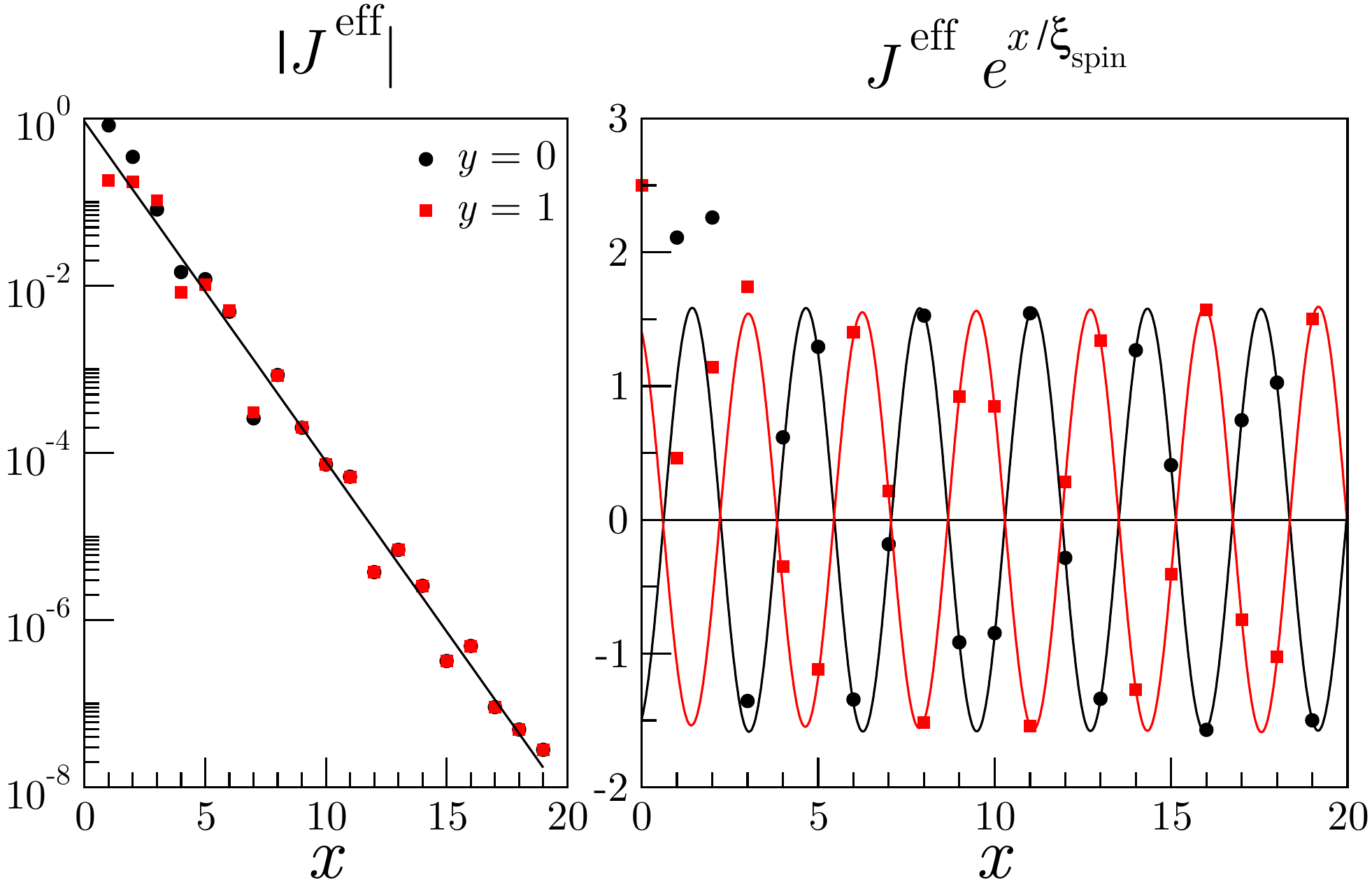}
\caption{(color online) Behavior of the effective interaction
  $\Jeff{x,y}$ between to impurities as a function of their relative
  distance for $J_\perp=3J_1$ and $J_2=J_1/2$. DMRG results (symbols)
  are fitted using the expression \eqref{eq:Jeff_fit}.}
\label{fig:Jeff}
\end{figure}

\begin{figure}[t]
\centering
\includegraphics[width=\linewidth,clip]{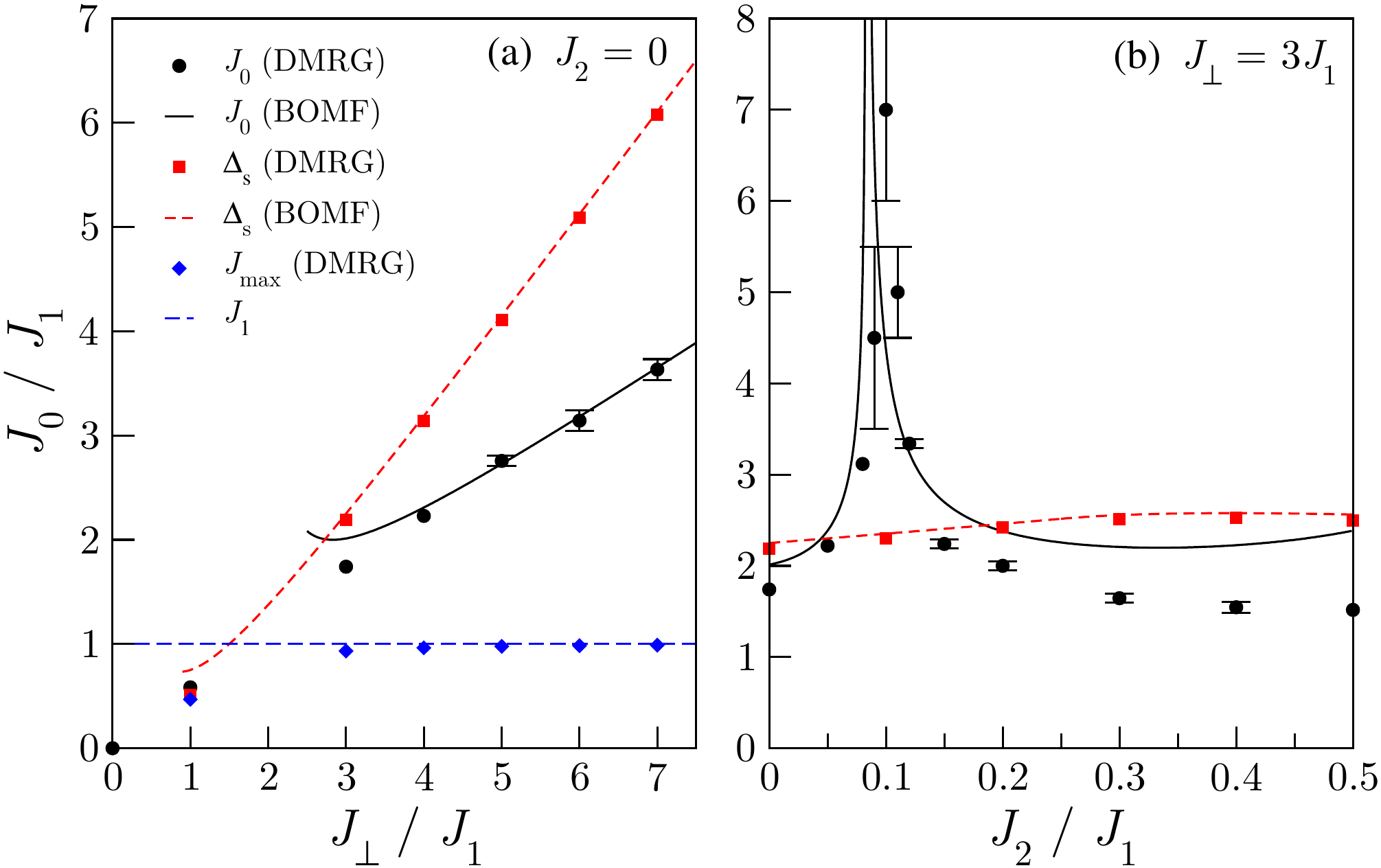}
\caption{Amplitude $J_0$ of the effective interaction (a) Evolution
  without frustration ($J_2=0$) for increasing coupling
  $J_\perp/J_1$. Comparison with other energy scales is given: the
  mean antiferro magnetic couplings $J_+$ and the maximum value of the
  effective interaction couplings $J_{\text{max}}$. (b) Evolution in
  the strong-coupling regime $J_\perp=3J_1$ for increasing frustration
  $J_2/J_1$, showing the divergence at the transition to the
  incommensurate regime.}
\label{fig:J0}
\end{figure}

Numerically, we compute the effective interaction using DMRG by
targeting the lowest energies in the singlet and triplet sectors. We
assume that the lowest triplet excitation is due to the interaction
between the two impurities (the spin gap is large enough in this
system) so that one can use the relation
\begin{equation}
J^\text{eff}=E_{S_\text{tot}=1}-E_{S_\text{tot}=0}\;.
\end{equation}
Only the total $S_\text{tot}^z$ is fixed in DMRG calculations. Thus,
since the triplet sector has a contribution for $S^z_\text{tot}=0$,
one accesses to the amplitude $|J^\text{eff}|$ from the energy
difference of the first two energies in this sector. In order to get
the sign of $J^\text{eff}$, one compares the obtained energies in
sector $S^z_\text{tot}=0$ with the lowest in sector
$S^z_\text{tot}=1$. On figure \ref{fig:Jeff}, we fit the curves using
the function
\begin{equation}
\label{eq:Jeff_fit}
\Jeff{x,y}=J_0 (-1)^{y+1} e^{-x/\xi_\text{spin}} \cos(qx+\phi)\;,
\end{equation}
where $q=\pi$ and $\phi=0$ in the commensurate regime.  As expected,
the wave-vector $q$ and length-scale $\xi_\text{spin}$ exactly
correspond to the ones of the magnetization profile and correlations
function. The behavior of the amplitude $J_0$ is not quantitatively
predicted by the BOMF theory as, for instance, the behavior of
$\xi_\text{spin}$ is not in perfect agreement, due to the
approximations made in the BOMF.  Still, as one sees on
figure~\ref{fig:J0}, that the amplitude $J_0$ displays a sharp
increase in the vicinity of the commensurate-incommensurate
transition, reminiscent from the divergence expected in
Eqs.~\eqref{eq:Jeff_commensurable}-\eqref{eq:Jeff_incommensurable}.

\subsubsection{Short distances behavior}

\begin{figure}[t]
\centering
\includegraphics[width=0.7\columnwidth,clip]{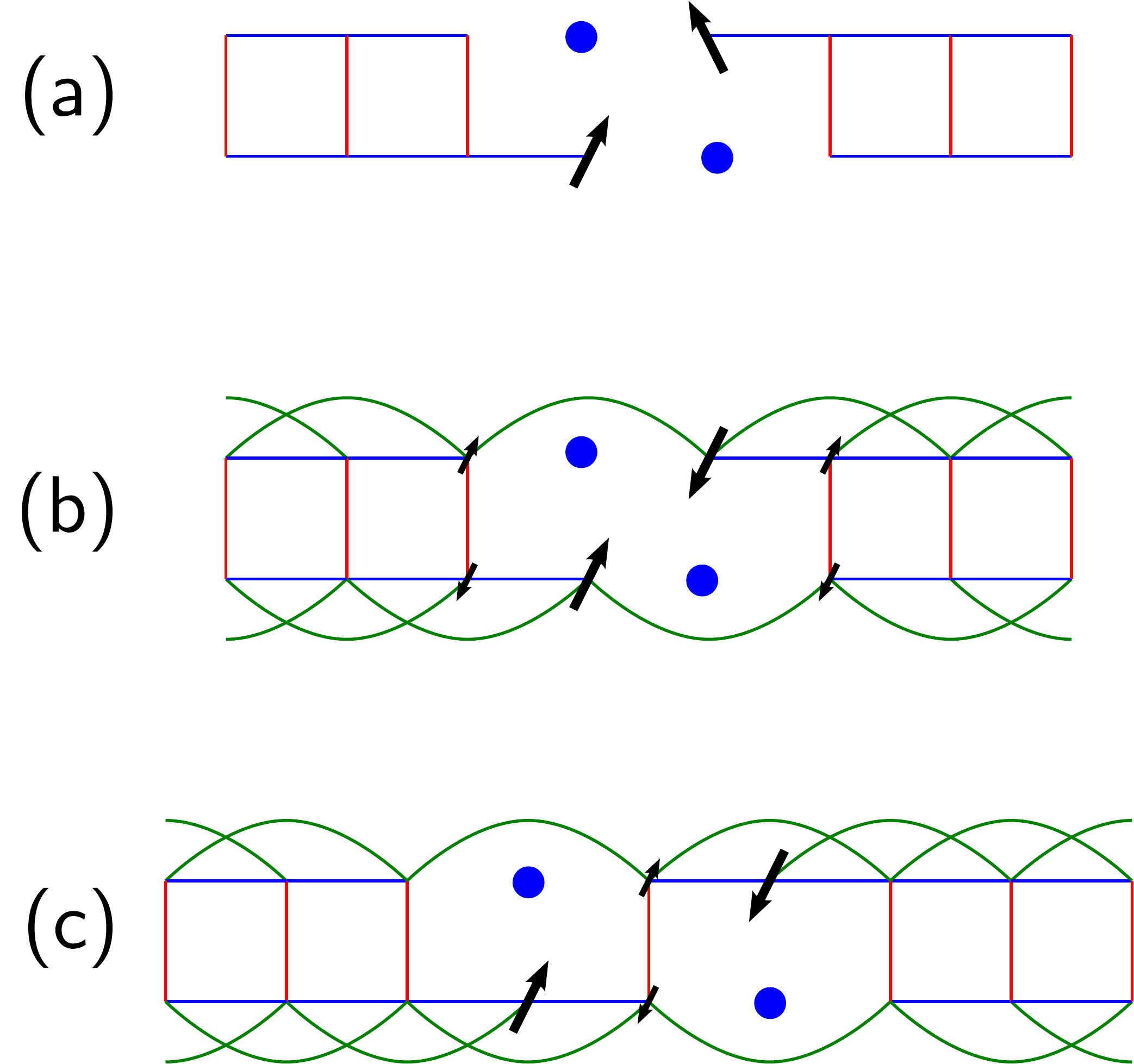}
\caption{(color online) Impurities configurations for which the
  effective interaction does not have the sign expected from 
  relation \eqref{eq:Jeff_fit}: (a) $\mathbf r=(1,1)$ with $J_2=0$,
  (b) $\mathbf r=(1,1)$ with $J_2\neq0$ and (c) $\mathbf r=(2,1)$ with
  $J_2\neq0$.}
\label{fig:interaction_courte_distance}
\end{figure}

The short distances effective interactions computed with DMRG, and
displayed on Fig.~\ref{fig:Jeff}, do not follow the prescription
\eqref{eq:Jeff_fit}. In fact, although linear response theory fails at
these distances, one can guess the sign and magnitude of $\Jeff{}$ by
looking at each configuration (see
Fig~\ref{fig:interaction_courte_distance}). The first thing to notice
is that a configuration with impurities on the same rung breaks the
ladder and the elementary excitation is then a magnon in the largest
piece which energy cost is slightly larger than the spin gap. Such
effect does not enter in the effective model since no spin-$1/2$ is
located at the vicinity of an impurity in that case.

Without frustration ($J_2=0$), the effective interaction oscillates at
$\mathbf k = (\pi,\pi)$, even at short distances, except when $\mathbf
r=(1,1)$ for which the effective coupling is almost zero within DMRG
accuracy.  Indeed, we observe on figure
\ref{fig:interaction_courte_distance}(a) that this configuration
breaks the ladder. Two spinons are then generated on disconnected
fragments and behave independently, making the triplet and singlet
states degenerate and the effective interaction equal to zero.  The
largest value of the effective interaction, which we write
$J_{\text{max}}$ in the following, is obtained when two impurities are
neighbour on the same chain. The magnitude is then controlled by $J_1$
and shown on Fig.~\ref{fig:J0}(a).

In the presence of frustration, the configuration with $\mathbf
r=(1,1)$ no longer breaks the ladder.
Figure~\ref{fig:interaction_courte_distance}(b) shows that spinons
freed by the impurities should anti-align due to $J_2$, as for the
configuration with $\mathbf r=(2,1)$ sketched on
\ref{fig:interaction_courte_distance}(c). In both cases, the
corresponding effective interaction is expected to be
antiferromagnetic (positive), in agreement with the DMRG result of
figure \ref{fig:Jeff}. If $J_2$ is larger than $J_1$, it typically
sets the scale of the maximum coupling $J_{\text{max}}$ in the
effective interaction.

Last, one can recall that two-body interactions is just an
approximation and that terms involving more than two partners should
be included to improve the comparison with ab-initio calculations
involving many impurities. The validity of two-body interaction has
been discussed in Ref.~\onlinecite{Mikeska1997} to which we refer to
for further details on this question.

\subsection{Distribution of couplings between impurities}

We here discuss the nature of the couplings distribution $P(J)$
resulting from doping the ladder and which is a central quantity for
the understanding of the magnetic responses. We use the following
notation from now on: $N$ is the total number of sites, $L=N/2$ the
length of the ladder, $N_i$ the number of impurities and $\doping =
N_i/N$ the impurities concentration or doping. The latter corresponds
to the probability for a site to be occupied by an impurity. The
lattice spacing is taken to be one in both directions. The relative
distance between two points on the ladder is written as $\mathbf{r} =
(n_x,n_y)$ with $n_x=0,1,2,\ldots$ and $n_y=0,1$.

Consider impurities that are randomly distributed on the ladder.  The
probability $p_{n_x,n_y}$ of having a distance $\mathbf{r}$ between
two impurities is given by a geometric law
\begin{equation}
p_{n_x,n_y} = \doping (1-\doping)^{2n_x+n_y-1} \quad \text{for}\quad n_x+n_y>0\;.
\end{equation}
To understand this formula, one can scan all intermediate sites
between the impurities following a zig-zag path. Thus, within the
ladder geometry which has the peculiarity to differ from a chain
because of the possibility to put two impurities on the same rung
(case with $n_x=0$ and $n_y=1$), we have the following results for the
mean longitudinal and transverse distances:
\begin{align}
\bar{x} &= \frac{1}{\doping}\left(\frac{1-\doping}{2-\doping}\right)\;, \\
\bar{y} &= \frac{1}{2-\doping}\;.
\end{align}
One recovers the intuitive behaviors $\bar{y}\simeq 1/2$ and $\bar{x}
= 1/(2\doping)$ in the dilute limit $\doping \ll 1$. Thus, in this
limit, the typical average distance $d$ between impurities, as if they
were on a chain, is given by the effective doping $\doping'=2\doping$
as $d \simeq 1/\doping'$ and one has to keep in mind the presence of
this factor two in qualitative reasoning.

To obtain the distribution of couplings, we use for analytical
calculations the simplified and generic relation
\begin{equation}
J(\mathbf{r}) = J_0 (-1)^{y+1} \cos(q x+\phi) e^{-x/\xi}
\end{equation}
with $q$ a dimensionless wave-vector which accounts for a possible
incommensurability and $\xi$ the spin correlation length (in a
shortened notation), $\phi$ a phase-shift and $J_0$ an energy
scale. Their typical behavior with microscopic parameters was
discussed in the previous subsections. Formally, one obtains the
distribution of couplings using the definition
\begin{equation}
p(J) = \iint d\mathbf{r}\; p(\mathbf{r}) \delta(J-J(\mathbf{r}))
\end{equation}
and use for the discrete case
\begin{equation}
p(\mathbf{r}) = \sum_{(n_x,n_y)} p_{n_x,n_y} \delta(x-n_x)\delta(y-n_y) \;.
\end{equation}
As we will see, the magnetic curve will be deeply connected to the
coupling repartition function that we denote by
\begin{equation}
R(J) = \int_{-\infty}^J P(J')dJ' \;.
\end{equation}
We also introduce the repartition function of antiferromagnetic
couplings only :
\begin{equation}
R_+(J) = \frac{R(J)-R(0)}{1-R(0)}\quad \text{for}\; J>0\;.
\end{equation}
Indeed, negative $J$s corresponding to ferromagnetic couplings will
yield polarized impurities as soon as the field is turn on.  A correct
way of defining an energy scale corresponding to a magnetic field in
the problem is thus to average only the positive $J$s. Then, we take
the following definition
\begin{equation}
J_{+} = \frac{\int_{0}^{\infty} J P(J)dJ}{\int_{0}^{\infty} P(J)dJ}
\end{equation}
for the typical energy scale of the antiferromagnetic couplings.

\subsubsection{Commensurate case}

In this case, the interaction is purely antiferromagnetic
corresponding to $q=\pi$. Changing variables is done using
\begin{equation*}
\delta(J-J(\mathbf{r})) = \frac{\xi}{|J|} \delta(x-n(J))\;,
\end{equation*}
with $n(J) = \xi \ln(J_0/|J|)$.

\paragraph{Continuous distribution --} We first consider the most
elementary situation where $p(\mathbf{r})$ is approximated by a
continuous function, which requires $\doping \ll 1$ and $\xi \gg 1$
with fixed $\doping \xi$. Then, what enters in the distance
probability is the effective chain doping $\doping'$, giving the
exponential law $p(x) \simeq \doping' e^{-\doping'x}$. The calculation
yields a symmetric power-law distribution
\begin{equation}
p(J) = \frac{\doping\xi}{J_0}\left(\frac{J_0}{|J|}\right)^{1-2\doping\xi}
\label{eq:PJ-powerlaw}
\end{equation}
for $J \in [-J_0,J_0]$, featuring the exponent $1-2\doping\xi$. The
corresponding repartition function reads
\begin{equation}
R(J) = \frac 1 2 \left[ 1 + \text{Sign}(J)\left(\frac{|J|}{J_0}\right)^{2\doping\xi} \right] \;.
\label{eq:rep-com}
\end{equation}
The energy scale $J_{+}$ takes the simple form
\begin{equation}
J_{+} = \frac{2\doping\xi}{1+2\doping\xi} J_0 \;.
\label{eq:meanJ-continuous}
\end{equation}
A similar expression has been used to interpret experiments with 3D
effects~\cite{Bobroff09}. Here, the effective volume of the
interaction boils down to $2\xi$.

\paragraph{Exact distribution and lattice effects} -- The continuous
distribution ansatz is not justified for systems with very short
correlation length such as the isotropic ladder. We here carry out the
calculation in the discrete case to obtain the exact formula that can
be compared to numerical histograms of the couplings used in numerical
simulations. We have
\begin{equation*}
p(J) = \sum_n P(J) \delta(n - n(J))\;,
\end{equation*}
and
\begin{equation*}
 P(J) = \frac{\doping \xi}{(1-\doping)J_0}\left(\frac{J_0}{|J|}\right)^{1+2\xi\ln(1-\doping)}
\end{equation*}
for $J \in [-J_0,J_0]$. One recovers the result of the continuous
approximation under its assumption. In particular, one can see that
lattice effects decouple the effect of the correlation length and of
the impurity concentration, i.e. the exponent is not a function of
$\doping\xi$ only, but a function of both $\doping$ and $\xi$, which
makes the results not so universal. Last, we notice that the
distribution becomes flat for the particular value
\begin{equation}
  z^* = 1 - e^{-1/(2\xi)}\;,
\label{eq:z-star}
\end{equation}
and we will see that this can have consequences on the shape of the
magnetic curve.

\begin{figure}[t]
\centering
\includegraphics[width=0.85\columnwidth,clip]{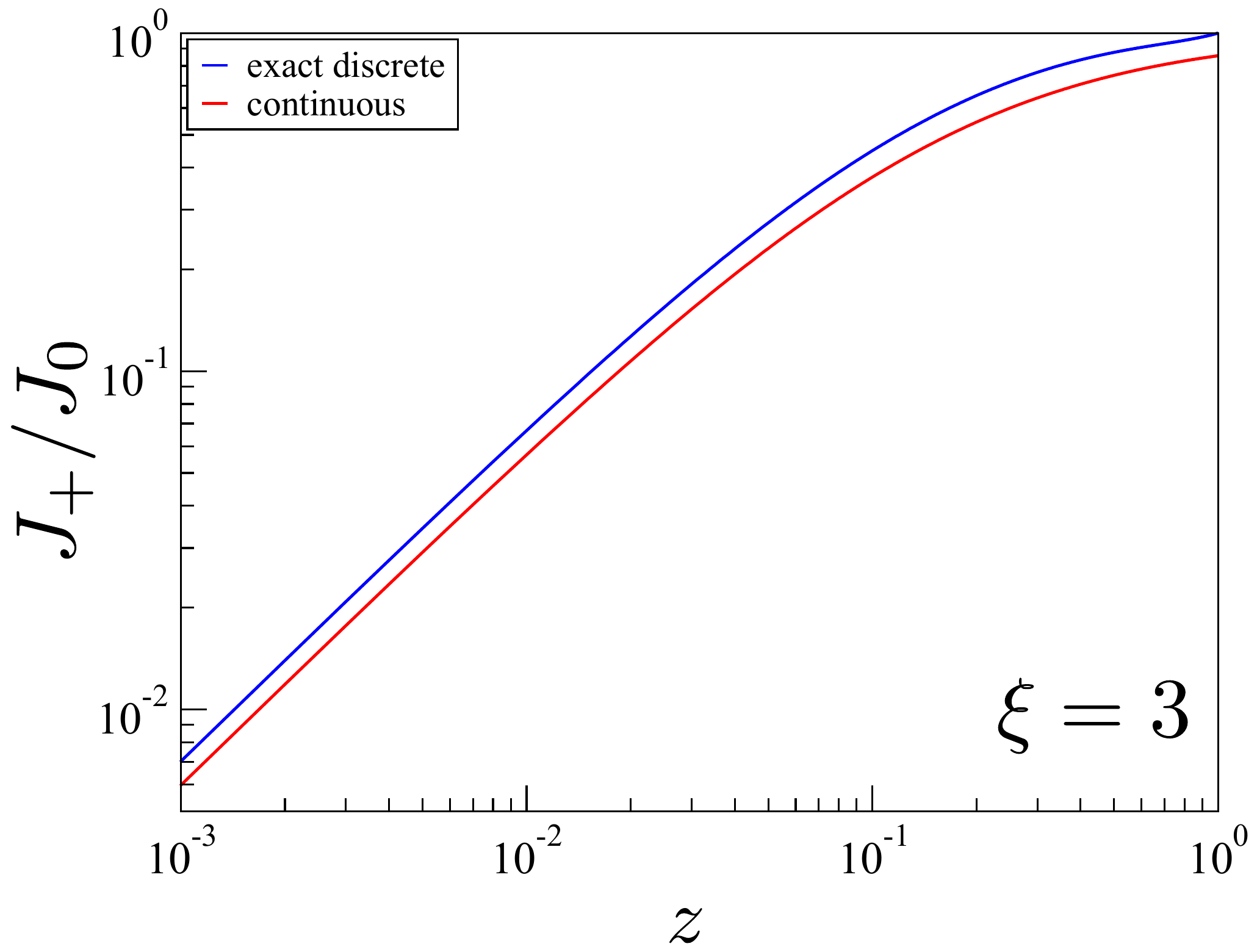}
\caption{(Color online) Typical energy scale of the coupling
  distribution as a function of doping $\doping$ for a realistic value
  of the correlation length $\xi = 3$. The continuous approximation
  result is compared to the exact discrete result, showing lattice
  effects discussed in the text.}
\label{fig:MeanJ}
\end{figure}

For the typical energy scale $J_+$, the exact result is computed by
directly summing upon the $(n_x,n_y)$ and one obtains
\begin{equation}
J_{+} = \frac{1-(1-\doping)^4}{2-\doping}\frac{1+(1-\doping)e^{-1/\xi}}{1-(1-\doping)^4e^{-2/\xi}} J_0\;.
\end{equation}
Here again, one can check that \eqref{eq:meanJ-continuous} is
recovered provided $\xi \gg 1$ and $\doping \ll 1$ while keeping $\xi
\doping$ finite. Otherwise, deviations from
\eqref{eq:meanJ-continuous} occur at all doping. In particular, the
low doping regime $\doping \ll 1$ at fixed $\xi$ is
\begin{equation}
\frac{J_{+}}{2J_0\doping} \simeq \frac{1+e^{-1/\xi}}{1-e^{-2/\xi}} = \xi\left(1-\frac{1}{2\xi}+\cdots\right)\;.
\end{equation}
These lattice effects are illustrated in Fig.~\ref{fig:MeanJ} for a
realistic case with $\xi=3$, which is characteristic of the isotropic
ladder limit.

\subsubsection{Incommensurate case and numerical sampling}

In this case the $J(\mathbf{r})$ function is not bijective which
changes qualitatively the distribution of the $J$s. The presence of
the cosine significantly lowers the weight of the largest $J$s while
the smallest $J$s will see their weight increase. Second, in the
presence of fractional $q/\pi$, commensurate effects happen while an
irrational $q/\pi$ has qualitatively the behavior of a true
quasi-periodic signal. In particular, for rational $q/\pi$ and
$\phi=0$, a fraction of couplings can be zero. Yet this situation is
unphysical for the model under consideration which generic case is
non-zero phase shift and irrational $q/\pi$.

\begin{figure}[t]
\centering
\includegraphics[width=\columnwidth,clip]{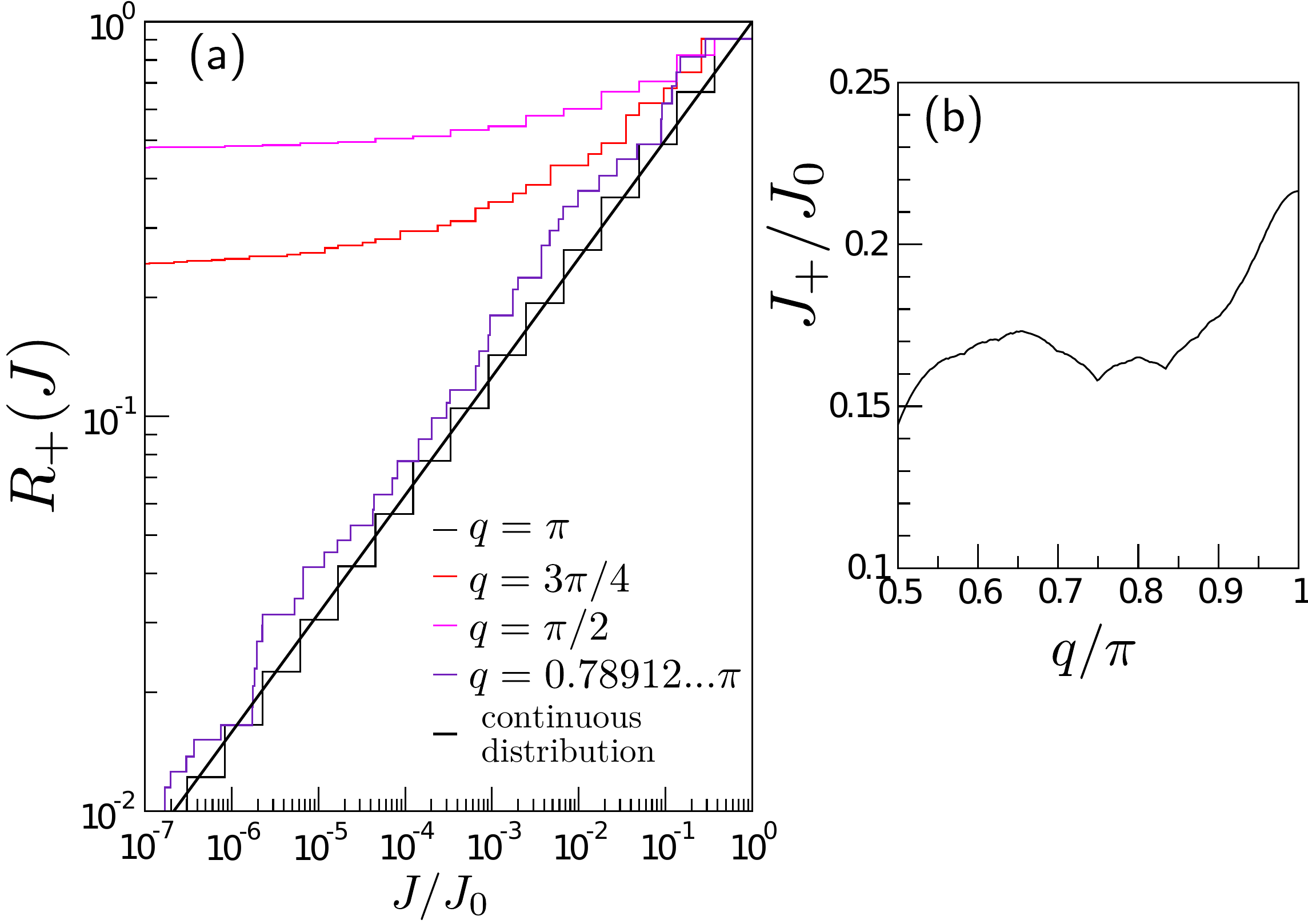}
\caption{(Color online) \textsf{(a)} Repartition function of
  positive couplings $R_+(J)$ for $\doping = 0.05$ and $\xi = 3$ with
  $\phi=0$. \textsf{(b)} energy scale $J_{+}$ vs the incommensurate
  wave-vector $q$.}
\label{fig:repartition}
\end{figure}

In order to illustrate the typical behavior of the repartition
function in the commensurate and incommensurate regimes, we have
sampled numerically the distribution of couplings. Results are
gathered in Fig.~\ref{fig:repartition}. The essential features are the
following: (i) up to discrete effect, the exponent $2\doping\xi$ in
Eq.~\eqref{eq:rep-com} captures well the power-law in the commensurate
case; (ii) for irrational $q/\pi$, the repartition function is
qualitatively very close to the commensurate case, with a similar
exponent, and up to the weight redistribution towards lower $J$ which
translates into a smaller energy scale $J_{+}$. (iii) For rational
$q/\pi$ and $\phi=0$, plateaus appear at $1/4$ and $1/5$ in the
figure, corresponding the many zero couplings, together with cusps in
$J_{+}$. Yet, the latter situation being unphysical, the main
conclusion is that incommensurability hardly affects the coupling
distribution. The fact that frustration, throughout
incommensurability, only lowers the energy scale $J_{+}$ but hardly
affects the distribution is essential to understand that frustration
will have only minor effects on the local magnetic responses studied
in the next sections.

\section{Exact zero-temperature results on a bipartite lattice:
  averaged spin and Curie constant}

In this section, we improve on the work of Sigrist and
Furusaki\cite{Sigrist1996} for the calculation of the averaged total
spin and zero temperature Curie constant of a doped system on a
bipartite lattice. The results that are obtained are more general than
for the special case of a ladder, and can be useful in checking
numerical simulations and understanding finite-size corrections.

\subsection{Total spin distribution}
\label{sec:CzNi}

We assume a finite size sample containing $N$ sites and doped with
$N_i=\doping N$ impurities, where $\doping$ is the impurity
concentration which is fixed. The impurities are assumed not to break
the lattice into disconnected sublattices (see discussion
Section~\ref{sec:chainbreak}). On a bipartite lattice, with two
sublattices $A$ and $B$ which have the same number of sites $N/2$,
applying Marshall's theorem yields that the total spin $S$ of a given
impurity configuration reads:
\begin{equation}
S=\frac 1 2 |N_{i,A}-N_{i,B}|=\frac 1 2 |2N_{i,A}-N_i|\;,
\end{equation}
where $N_{i,A}$ (resp. $N_{i,B}$) is the number of impurities on
sublattice $A$ (resp. $B$). The probability of having a configuration
with $N_{i,A}$ impurities on sublattice $A$, is qualitatively similar
to the result on a ferro-antiferromagnetic (F-AF)
chain~\cite{Frischmuth1999}, and given by
\begin{equation}
P(N_{i,A}) =\displaystyle \frac{\binom{N/2}{N_{i,A}}\binom{N/2}{N_i-N_{i,A}}}{\binom{N}{N_i}}\;.
\end{equation}
Then, the probability of having a total spin $S$ on a sample is
\begin{equation}
P_{\doping,N_i}(S)=\displaystyle \frac{\binom{N/2}{\frac {N_i} 2 + S}\binom{N/2}{\frac {N_i} 2 - S}}{\binom{N}{N_i}}(2-\delta_{S,0})\;,
\end{equation}
where $S \in [0,N_i/2]$.  This result is exact and can be used to
compute numerically the mean total spin and the Curie constant. For
large $N_i$ and fixed $\doping$, according to the central limit
theorem, $P_{\doping,N_i}(S)$ converges towards a Gaussian. A
saddle-point calculation gives the asymptotic behavior:
\begin{equation}
P_{\doping,N_i}(S) \simeq \frac{2}{\sqrt{2\pi\sigma_S^2}} e^{-S^2/2\sigma_S^2}\;,
\end{equation}
of variance $\displaystyle \sigma_S^2 = \frac{N_i}{4}(1-\doping)$.

\subsection{Total spin and Curie constant}

\begin{figure}[t]
\centering
\includegraphics[width=\columnwidth,clip]{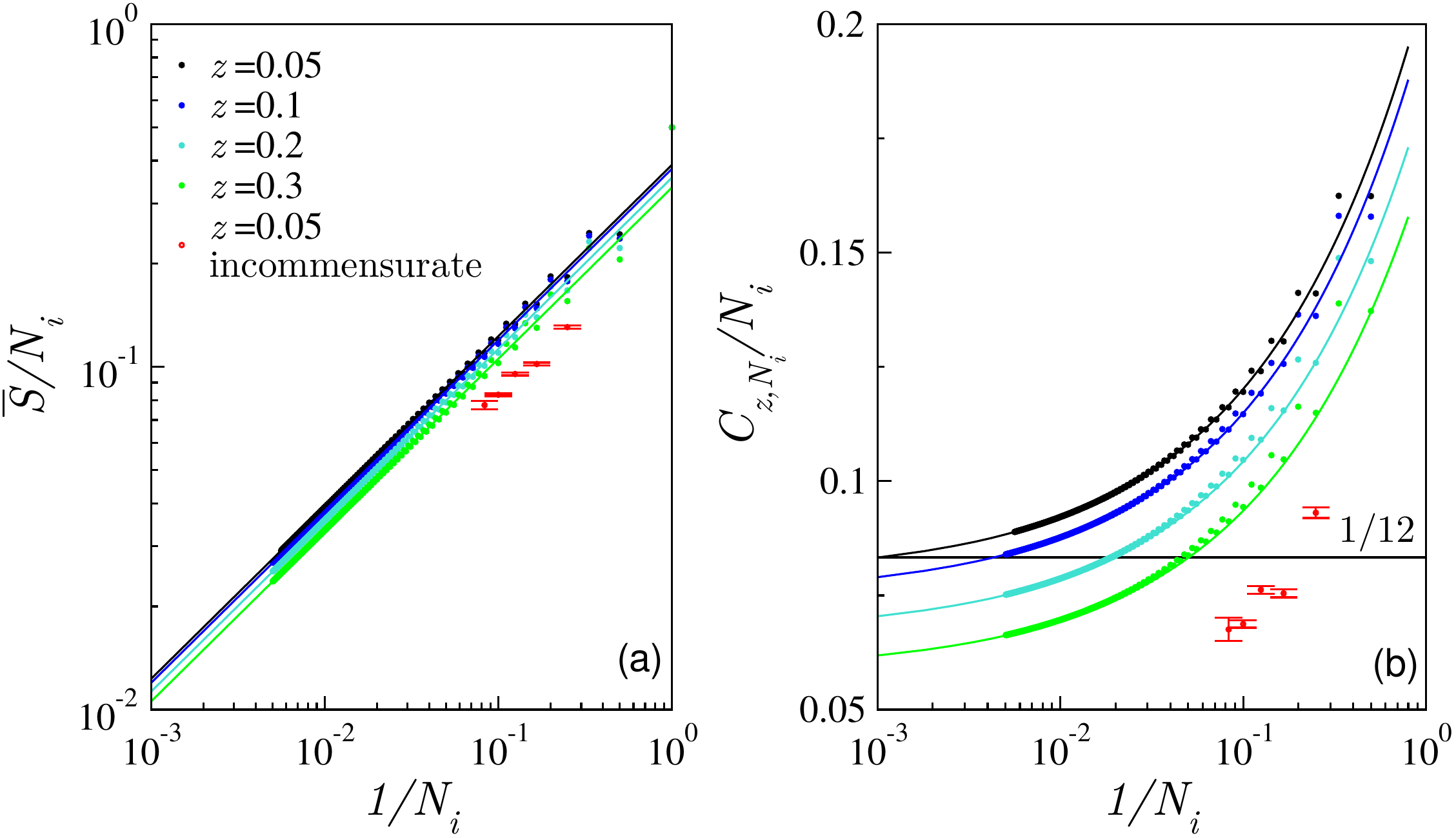}
\caption{(Color online) \textsf{(a)} Finite-size scaling of the
  average total spin $\overline{S}$, for four different doping
  $\doping$, obtained from exact calculations (points) and compared to
  Eq.~\eqref{eq:Smoy}. \textsf{(b)} The finite-size correction of the
  zero-temperature Curie constant, compared to
  Eq.~\ref{eq:c-exact-ladder}. The $1/12$ line shows the usually
  admitted result. The incommensurate case corresponds to an isotropic
  ladder with $J_2=0.8J_1$.}
\label{fig:finite-size-C}
\end{figure}

One then obtains the average spin and the average square-spin in the
$N\rightarrow\infty$ limit as
\begin{equation}
\overline{S}\simeq\sqrt{\frac{1-z}{2\pi}}\sqrt{N_i}\quad\text{and}\quad\overline{S^2}\simeq \frac{1-z}{4}N_i\;.
\label{eq:Smoy}
\end{equation}
The total zero-temperature Curie constant matches exactly
\begin{equation}
C_{\doping,N_i} = \overline{\moy{\hat{S}_z^2}} = \frac{\overline{S(S+1)}}{3}\;.
\end{equation}
From \eqref{eq:Smoy}, we obtain the following asymptotic behavior
\begin{equation}
C_{\doping,N_i} = \frac{N_i}{12}(1-\doping)\left[1 + \sqrt{\frac{8}{\pi(1-\doping)}}N_i^{-1/2}\right]\;.
\label{eq:c-exact-ladder}
\end{equation}
These asymptotic results are compared to numerical calculations using
the exact distribution on Fig.~\ref{fig:finite-size-C}. In the
thermodynamical limit $N_i \gg 1$, we thus find that the Curie
constant \emph{per impurity} is $(1-\doping)/12$ at $T=0$, while the
Curie constant \emph{per spin} is $\doping/12$.  Notice that one can
also compute exactly the average of the squared spin:
\begin{align*}
\overline{S^2} = \Bigg[\frac{N_i^2}{4}\binom{N}{N_i} & -N(N_i-1)\binom{N-1}{N_i-1} \\
 &+ N(N-1)\binom{N-2}{N_i-2} \Bigg]/\binom{N}{N_i}\;,
\end{align*}
which is useful to cross-check the statistical convergence of
averaging over samples, but we did not manage to compute exactly
$\overline{S}$.

These results remain correct in the commensurate regime because the
effective model is still unfrustrated. But in the incommensurate
regime, expressions \eqref{eq:Smoy} are not valid anymore. The total
spin and Curie constant at zero-temperature can be computed from exact
diagonalization on the effective model using the exact couplings
extracted from DMRG data. The results for a frustrated isotropic
ladder in the incommensurate regime are shown on
Fig.~\ref{fig:finite-size-C}. The frustration induced by the
incommensurability yields a appreciable reduction of both the total
spin and the Curie constant. Note that this reduction is essentially
due to the short distance behavior of the effective interaction, hence
the necessity to use the exact effective couplings computed in DMRG
rather than the asymptotic law \eqref{eq:Jeff_fit}.

The doping dependence of the prefactors and finite-size corrections
were missed in previous work. They originate from the dilution of the
lattice. They actually play a crucial role in the quantitative
understanding of the numerics that usually work with a restricted
number of impurities. Last, the precise value at finite-size is
essential in extracting the low-temperature exponent, as we will see
in the next Section.

\subsection{Chain breaking effects on the ladder}
\label{sec:chainbreak}

\subsubsection{Percolation scenario in the general case}

Results in Eqs.~\eqref{eq:Smoy} and \eqref{eq:c-exact-ladder} are
exact assuming that there is a fully connected cluster containing
$N_i$ impurities over $N=N_i/\doping$ sites. Of course, the network
can be disconnected in many sub-clusters by the presence of
impurities. Then, each cluster, which is a finite size system, will
eventually contribute to $\bar{S}$ and $C$ in the thermodynamical
limit, which was pointed out by Sigrist and Furusaki who computed an
evaluation of the correction in the ladder case
\cite{Sigrist1996}. Thus, we expect that, in general, $\bar{S}/N_i
\neq 0$ in the thermodynamical and that $C$ is not exactly given by
\eqref{eq:c-exact-ladder}. Lastly, it has been proposed in
Ref.~\onlinecite{Sigrist1996} that these breaks provide an natural cut
in the length scale (and an energy scale) which should affect the
behavior of the correlations. This was confirmed numerically in
Ref.~\onlinecite{Trinh2012} in which results on depleted two-leg
ladders are consistent with an upper bound of the order of $z^{-2}$
reached for very low temperatures.

The probability and size distribution of these clusters are governed
by percolation theory. The percolation transition distinguishes two
main regimes: (i) there exist an infinite cluster of sites below a
certain critical doping $\doping_c$, (ii) for $\doping>\doping_c$,
only finite-size clusters exist and their size distribution is
typically exponential, associated with a mean cluster size that will
be denoted by $\bar{\ell}$ in the following. At the critical doping
$\doping=\doping_c$, there is still an infinite cluster and scaling is
expected for the mean cluster size. The value of $\doping_c$ is very
sensitive to dimensionality and connectivity of the lattice. These
percolation regimes induce important finite size effects and are
essential for experiments and numerical simulations. On a chain, it is
clear that $\doping_c=0$, i.e. any finite doping will break the
lattice and the mean cluster size is easily related to the doping
$\bar\ell \sim 1/\doping$. On a ladder, the situation is similar in
the sense that any finite doping breaks the lattice into
sub-clusters. This chain breaking effects have been shown to have
quantitative results on the magnetic response of doped chains
\cite{Sirker2007}. Yet, computing $\bar{\ell}$ in the case of ladder
is not trivial and the remaining clusters are themselves doped with
various concentration of impurities which makes the predictions more
involved. We give below an exact discussion of the cluster sizes in
the ladder and apply the results to the chain breaking effects on
ladders.

\subsubsection{Cluster distribution for the unfrustrated ladder network}

We consider a ladder with nearest-neighbour only. Connectivity of the
network is broken if (i) two impurities fall on the same rung, or (ii)
two impurities fall on diagonal positions on a plaquette (see figure
\ref{fig:interaction_courte_distance}(a)).  If $(x,0)$ is the impurity
position, there are three positions at which a second impurity can
break the ladder: $(x-1,1)$, $(x,1)$ et $(x+1,1)$. Occupying a site
with an impurity has a probability $\doping$.  In the diluted limit
$\doping \ll 1$, the density of cuts is then $3\doping^2$ and the
corresponding mean cluster size is given by $\bar{\ell} \simeq
1/3\doping^2$ (the factor 3 was missing in
Ref.~\onlinecite{Sigrist1996}). Notice that in the presence of
frustration, chain-breaking requires at least four neighboring
impurities which would make a different scaling $\bar{\ell} \sim
1/\doping^4$. For large enough distances, breaks are uncorrelated so a
fair description of the distribution law is that of a poissonian
process
\begin{equation}
\label{eq:rho_ell_exp}
\rho(\ell)\simeq\zeta e^{-\zeta\ell}\;,
\end{equation}
with $\zeta = 1/\bar{\ell} \simeq 3z^2$ in the diluted and continuum
limit.

An exact calculation of the cluster sizes distribution is carried out
in Appendix~\ref{app:breaks} and supports this phenomenological
approach. The exact distribution reaches very quickly an asymptotic
behaviour given by a geometric law
\begin{equation}
\label{eq:rho_ell_geo}
\rho(\ell)\simeq\zeta(1-\zeta)^{\ell-1}\;,
\end{equation}
of parameter
\begin{equation}
\zeta = \frac{1}{2}\left(1+z-(1-z)\sqrt{1+4z(1-z)}\right) \underset{z\ll1}{\simeq}3z^2\;.
\end{equation}
Consequently, one recovers \eqref{eq:rho_ell_exp} in the continuum
limit and $\bar{\ell}=1/\zeta$.  In particular, this provides
finite-$\doping$ corrections to the $1/\doping^2$ scaling which turn
out to be quantitative even for a few percent doping as we see now.

\subsubsection{Consequences for the averaged spin and Curie constant}

Averaging the total spin and Curie constant over clusters is not
trivial since the doping of each cluster can now be distributed
between zero and approximately $1/2$. To handle a correct estimate,
one would have to average using the joint distribution of cluster
sizes and doping. We give below a rough estimate that consists in
assuming a fix doping $\doping$ for all cluster and averaging only
over cluster sizes $\ell$ using $N_i = 2\ell \doping$. Neglecting the
doping fluctuations should yield a good approximation in the diluted
limit where cluster sizes diverge. Averaging the equations
\eqref{eq:Smoy} is performed using
\begin{equation*}
\overline{\overline{S_\text{tot}}}\simeq\sum_{\ell=1}^{+\infty}\rho(\ell)\sqrt{\frac{z}{\pi}(1-z)\ell}
\;\;\text{and}\;\;
\overline{\overline{S_\text{tot}^2}}\simeq \sum_{\ell=1}^{+\infty}\rho(\ell)\frac{z}{2}(1-z)\ell\;.
\end{equation*}
Using the approximate law \eqref{eq:rho_ell_geo} for the size
distribution and a continuous approximation for the average spin, we
get
\begin{equation}
\overline{\overline{S_\text{tot}}}\simeq \frac12 \sqrt{\frac z\zeta (1-z)}
\quad\text{and}\quad
\overline{\overline{S_\text{tot}^2}}\simeq \frac{z}{2\zeta}(1-z)(1-\zeta)\;.
\end{equation}
To obtain the density of spin and Curie constant, one has to multiply
them by the clusters density $\zeta$. The mean spin density
$\overline{\overline{s}} =
\overline{\overline{S_\text{tot}}}/\bar{\ell}$ and Curie constant
density $c =
\overline{\overline{S_\text{tot}(S_\text{tot}+1)}}/3\bar{\ell}$ now
read
\begin{equation}
\overline{\overline{s}} \simeq \frac12 \sqrt{z(1-z)\zeta}
\label{eq:s-density}
\end{equation}
and
\begin{equation}
c\simeq \frac{z}{6}(1-z)(1-\zeta)+\frac16 \sqrt{z(1-z)\zeta}\;,
\label{eq:c-density}
\end{equation}
which gives in the diluted limit $z\ll1$
\begin{equation}
\overline{\overline{s}} \simeq \frac{\sqrt3}2 z^{3/2}
\quad\text{and}\quad
c\simeq \frac{z}{6}(1+\sqrt{3\doping}+\ldots)\;,
\end{equation}
in agreement with the results of Sigrist and
Furusaki~\cite{Sigrist1996} up to a prefactor.

In the opposite limit of high-density for impurities $\doping
\rightarrow 1$, the system is equivalent to a few independent
spin-$1/2$s which essentially behave as in a paramagnetic
phase. Therefore, we have that $\overline{\overline{s}} \simeq
1-\doping$ and $c \simeq (1-\doping)/2$. In particular, we infer that
there is an optimal doping $\doping^{o}$ which maximizes the Curie
constant and a slightly different one which maximizes the total
spin. The value of $\doping^{o}$ is non-trivial since it occurs at the
crossing of the two asymptotes.

\begin{figure}[t]
\centering
\includegraphics[width=\columnwidth,clip]{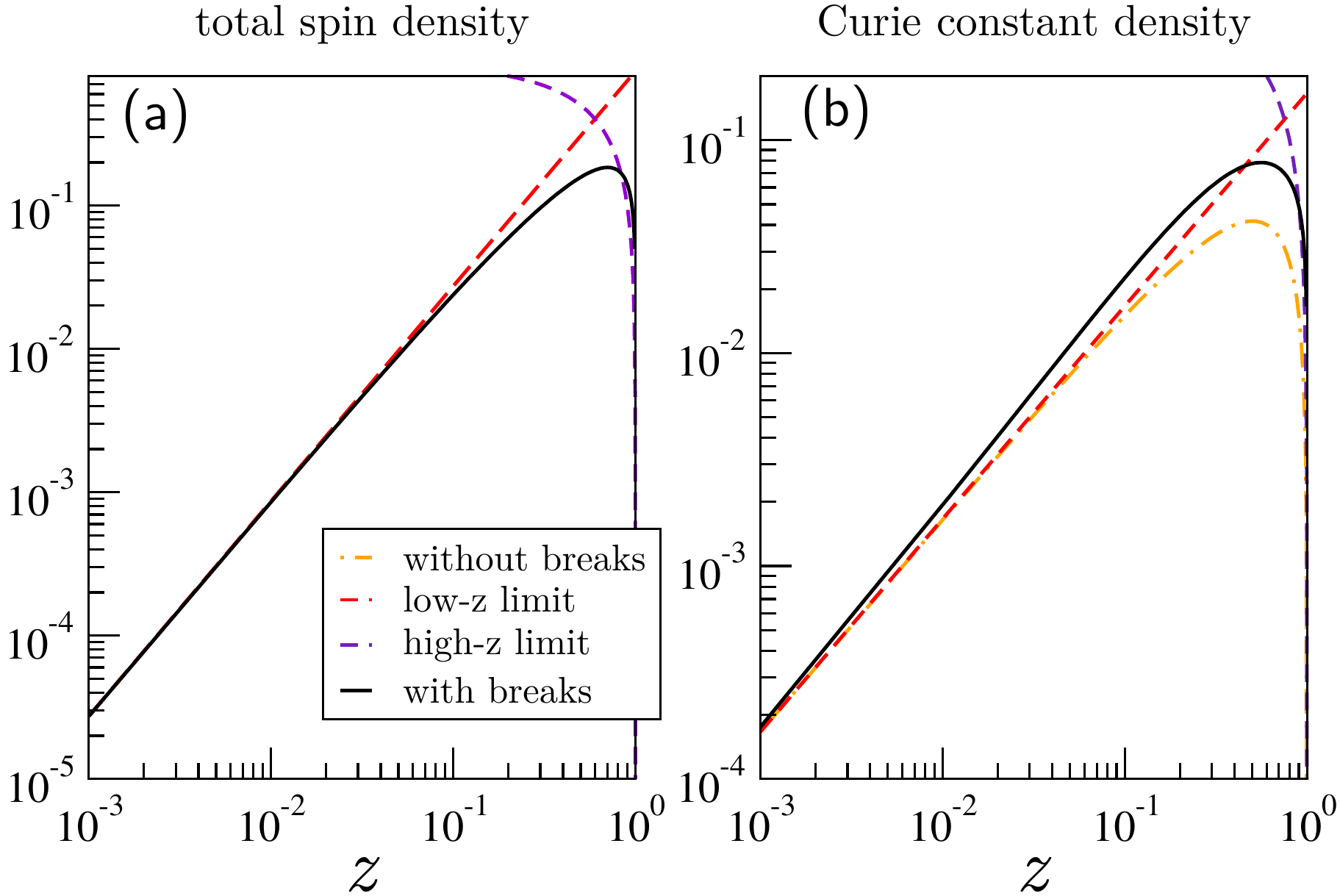}
\caption{(Color online) Effect of chain breaking on ladders (a) Mean
  total spin density $\overline{\overline{s}}$ and (b) Curie constant
  density $c$ averaged.}
\label{fig:chainbreaking}
\end{figure}

These predictions are a lower bound for $\overline{\overline{s}}$ and
$c$ since very small clusters with a few sites have a total spin and
susceptibility larger than the random-walk result. In order to show
the quantitative role of chain breaking and the validity of the
results, we plot on Fig.~\ref{fig:chainbreaking} the limiting
behaviors at low and high doping $\doping$ together with the formulas
\eqref{eq:s-density} and \eqref{eq:c-density}. Although these formula
are not exact, they capture well the existence of a maximum at an
optimal doping.

\section{Models and methods}

\subsection{Magnetization and Curie constant densities}

We now turn to the effect of interactions on the magnetic curve of
two-leg ladders doped with impurities. The magnetic excitation is
denoted by $H$. We choose to define the total magnetization density as
$m=\moy{\op{S}_z}/L = 2\moy{\op{S}_z}/N$ so that the high field
saturation density is $m_{sat} = 1-\doping$. With this normalization,
the contribution of non-interacting impurities carrying a spin-$1/2$
to the magnetization is the Brillouin formula at temperature $T$ (we
set $k_B=1$ in the following)
\begin{equation}
m(H,T) = \doping \tanh[H/2T]
\label{eq:brillouin}
\end{equation}
where $\doping$ thus corresponds to the \emph{impurity} saturation
magnetization, assuming that each impurity frees a spin-$1/2$. The
latter assumption is actually affected by chain breaking and will be
discussed in more details in Sec.~\ref{sec:magnetization}.

In the low-field limit at finite-temperature, the Curie constant
density $c$ is defined as
\begin{equation}
c = T\frac{\partial m}{\partial H}\bigg\vert_{H=0}\;.
\end{equation}
In the case of independent impurities, we would have the total Curie
constant $C_{\doping,N_i} = N_i/4$ corresponding to
$c=C_{\doping,N_i}/(N/2) = \doping/2$, as expected from
\eqref{eq:brillouin}. If one uses the exact result
\eqref{eq:c-exact-ladder} for correlated impurities, then the Curie
constant density reads $c=\doping(1-\doping)/6$ without chain breaking
or \eqref{eq:c-density} with it.

\subsection{Random dimer model}

We first consider a simple model, dubbed ``random dimer model'' in
which impurities are assumed to build dimers with their
neighbour. Dimers are independent but have couplings randomly
distributed according to $P(J)$.  The magnetization of a single dimer
of coupling $J$ is given by
\begin{equation}
m_{\text{dimer}}(H,T;J) = \frac{2\sinh[H/T]}{1 + e^{J/T} + 2 \cosh[H/T]}\;.
\end{equation}
The total number of dimers is $N_i/2$ so that the total magnetization
density averaged over the coupling distribution reads:
\begin{equation}
m(H,T) = 2\doping \int dJ P(J) \frac{\sinh[H/T]}{1 + e^{J/T} + 2 \cosh[H/T]} \;.
\label{eq:RDM-magnetization}
\end{equation}
The Brillouin formula \eqref{eq:brillouin} is recovered when
$P(J)=\delta(J)$ or more physically, in the high-temperature limit
when $T\gg J_{\text{max}}$.

Taking the zero-field limit at finite temperature in
\eqref{eq:RDM-magnetization} yields a temperature-dependent Curie
constant $c(T)$:
\begin{equation}
c(T) = 2 \doping \int dJ P(J) \frac{1}{3 + e^{\beta J}}
\label{eq:RDM-curie}
\end{equation}
which reaches the free spins result $c(T) = \doping/2$ in the
high-temperature regime.

\subsection{Solving the effective model}

The effective Hamiltonian of interacting impurity spins is given by
\begin{equation}
\op{\Ham}_\text{imp}= \sum_{\mathbf{I},\mathbf{J}} J_{\text{eff}}(\mathbf{I}-\mathbf{J}) \opspin{\mathbf{I}}\cdot\opspin{\mathbf{J}}
\label{eq:Himp-ED}
\end{equation}
and is solved numerically using either ED for $N_i =10$ or QMC up to
$N_i=100$ and provided there is no incommensurability, i.e. for
$q=\pi$. ED provides all energies $\{E_n(S_z)\}$ in a sector of total
spin $S_z$ so finite-temperature predictions are accessible. The
couplings $J_{\text{eff}}(\mathbf{I}-\mathbf{J})$ are obtained by
sampling impurities configurations on a ladder and using either the
approximate formula \eqref{eq:Jeff_fit} with chosen $q$, $J_0$ and
$\xi$, or the exact couplings computed from DMRG.

\subsection{Ab-initio calculations}

Two ``ab-initio'' methods are also used to compute observables
directly on the original microscopic Hamiltonian~\eqref{eq:microH}:
the DMRG technique, which gives accurate results for the
zero-temperature magnetization curve, and quantum Monte-Carlo SSE
calculations well suited for finite-temperature dependence.

\section{Zero-field susceptibility and the temperature-dependent Curie constant}

In this section, we focus on the limit of vanishing magnetic
excitation $H\rightarrow 0$ at finite $T$. The order of limits matters
and the situation $T\rightarrow 0$ for fixed $H$ will be studied in
the next Section. In this limit, a modified Curie law is generically
expected, written as
\begin{equation}
m(H,T) = \frac{c(T)}{T} H
\end{equation}
where $c(T)$ is a temperature-dependent Curie constant, corresponding
to a static susceptibility $\chi(T) = c(T)/T$. The goal of this
section is to investigate quantitatively the whole $c(T)$ curve and
analyze the effect of interactions, doping and frustration on its
behavior. The various regimes of $c(T)$ in depleted ladders were first
sketched by Sigrist and Furusaki~\cite{Sigrist1996} who gave the
following picture: starting at high-temperatures, the spins are
essentially independent because of thermal fluctuations so that $c(T)
\simeq (1-\doping)/2$. Assuming that the spin gap $\Delta_s$ is larger
enough than the maximum coupling $J_{\text{max}}$ (implicitly
corresponding to the strong-coupling regime), lowering the temperature
below the spin gap freezes all magnon excitations. Only remain
spin-$1/2$s freed by impurities which should behave independently for
a range of temperatures $J_{\text{max}} \lesssim T \lesssim
\Delta_s$. This yields a plateau around $c \simeq \doping/2$ if one
neglects chain breaking and $c \simeq \doping(1-\doping)/2$ if they
are taken into account to first order corrections.  Lowering again
temperature enables one to reach the zero-temperature plateau
discussed above and which is approximately given by $c \simeq \doping
/6$ (the $\doping/12$ plateau within Ref.~\onlinecite{Sigrist1996}
conventions). In the regime governed by impurity-spins interactions,
real-space renormalization group (RSRG) arguments\cite{Frischmuth1997}
generally gives low-temperature corrections of the form
\begin{equation}
c(T) \simeq c(0) + K (T/J_{\text{max}})^\alpha\;,
\end{equation}
with $K$ a non-universal constant and $\alpha$ an exponent which
generally depends on doping $\doping$ and that captures the
interesting physics about impurities interactions. We now check and
analyze this scenario using our various models and methods.

\subsection{Hints from the random dimer model}

\begin{figure}[t]
\centering
\includegraphics[width=0.95\columnwidth,clip]{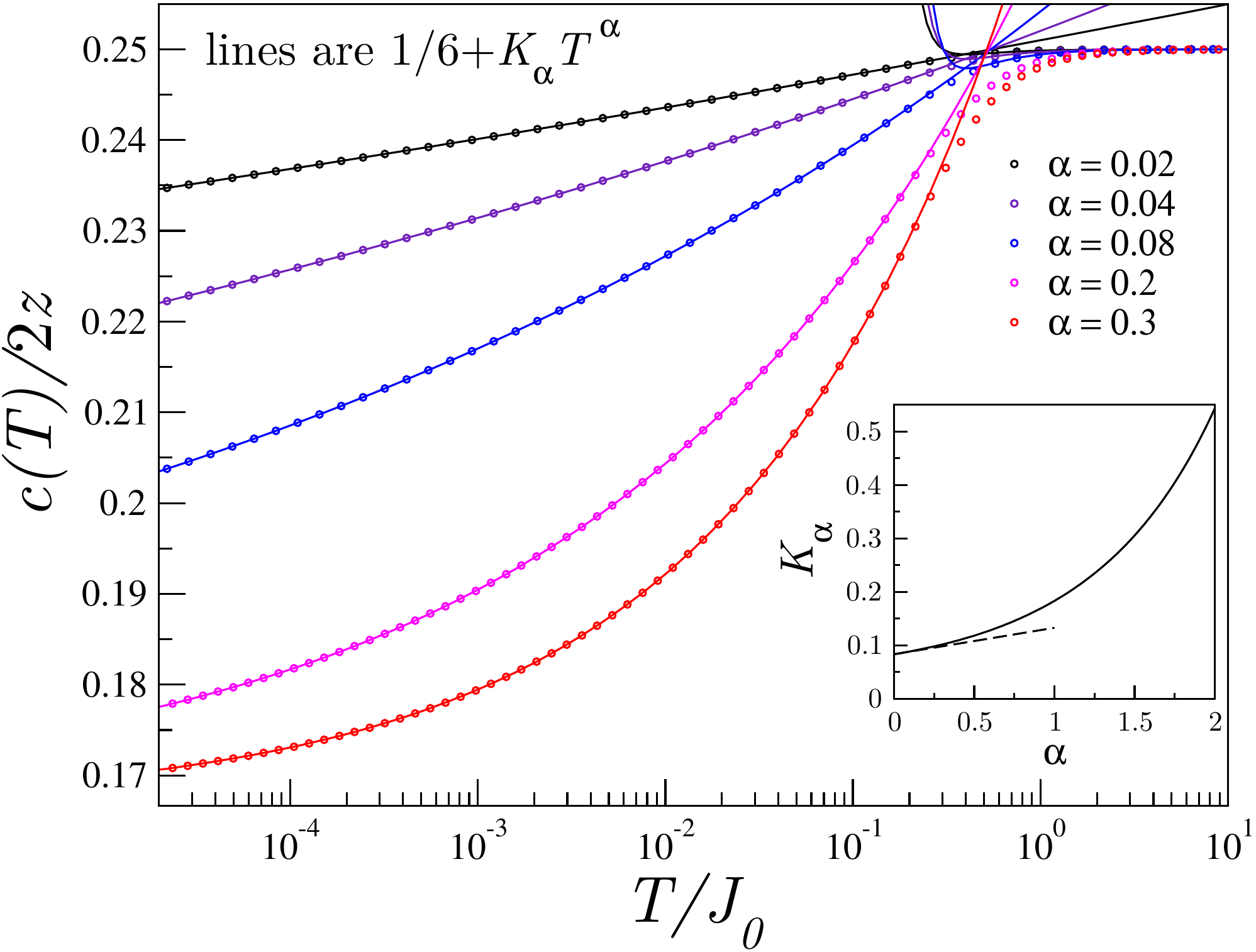}
\caption{(Color online) Curie constant within the random dimer
  model. In the ladder model $\alpha=2\doping\xi$. \emph{Inset}:
  behavior of the constant $K_\alpha$ from \eqref{eq:RDM-Kalpha}.}
\label{fig:dimermodel}
\end{figure}

The random dimer model formula \eqref{eq:RDM-curie} for the Curie
constant already displays a non-trivial temperature dependence due to
the coupling distribution. In the case of the power-law distribution
\eqref{eq:PJ-powerlaw} for which the exponent $\alpha = 2\doping\xi$
is the one of the couplings distribution and $J_{\text{max}}=J_0$ (see
below Eq.~\eqref{eq:RDM-cofT}), the high-temperature expansion leads
to
\begin{align*}
c(T) &= \frac \doping 2\left[1  -\frac{1}{16}\frac{\alpha}{\alpha+2}\left(\frac{J_0}{T}\right)^{2} \right.\\
& \left.+\frac{5}{768}\frac{\alpha}{\alpha+4}\left(\frac{J_0}{T}\right)^{4}
 -\frac{77}{184320}\frac{\alpha}{\alpha+6}\left(\frac{J_0}{T}\right)^{6} +\ldots \right]
\end{align*}
while, at low-$T$, a Sommerfeld-like expansion, in which the constant
three appearing the denominator of \eqref{eq:RDM-curie} has to be
carefully taken into account, yields a power-law:
\begin{equation}
\frac{c(T)}{2\doping} = \frac 1 6 + K_{\alpha} \left(\frac{T}{J_0}\right)^{\alpha}
\label{eq:RDM-cofT}
\end{equation}
with the constant 
\begin{equation}
K_{\alpha} = \frac 4 9 \int_0^1 du (-\ln{u})^{\alpha} \frac{1-u^2}{(1+\frac{10}{3}u+u^2)^2} \;.
\label{eq:RDM-Kalpha}
\end{equation}
$K_{\alpha}$ is of the order of $0.1$-$0.2$ and matches some simple
numbers  for specific values of $\alpha$: $K_0 = 1/12 \simeq 0.083333\ldots$, $K_1 = \frac{\ln 3}{6}
\simeq 0.183102\ldots$. The curves for
various $\alpha$ are represented in Fig.~\ref{fig:dimermodel} and show
that the prediction \eqref{eq:RDM-cofT} works for a wide range of
temperatures. In the limit of small $\alpha$, one has the expansion
$K_{\alpha} = 1/12 + K' \alpha$ with $K' = \frac 4 9 \int_0^1 du
\ln(-\ln{u}) \frac{1-u^2}{(1+\frac{10}{3}u+u^2)^2} \simeq
0.0493662\ldots$.

It is clear that, even though there is a power-law, there is nothing
universal in this result. The scaling originates only from the fact
that the distribution is a power-law. Last, the zero-temperature
result $c=\doping/3$, which already differs from free spins, is always
expected in the case of a symmetric $J \rightarrow -J$ distribution.

\subsection{Results on the effective model and a possible scenario for the low-temperature exponent}

In ab-initio ED and QMC calculations, the temperature-dependent Curie
constant is computed exactly using the average over thermal states and
disorder configurations:
\begin{equation}
c(T) = \frac{2}{N}\overline{\frac{\sum_{S_z}\sum_n S_z^2 e^{-E_n(S_z)/T}}{\sum_{S_z}\sum_n e^{-E_n(S_z)/T}}}\;,
\end{equation}
since $\moy{S_z}=0$ when $H=0$ for both the microscopic and effective
models due to SU(2) symmetry. Notice that, on the effective model,
chain breaking effects discussed in Sec.~\ref{sec:chainbreak} are not
included.

\begin{figure}[t]
\centering
\includegraphics[width=0.95\columnwidth,clip]{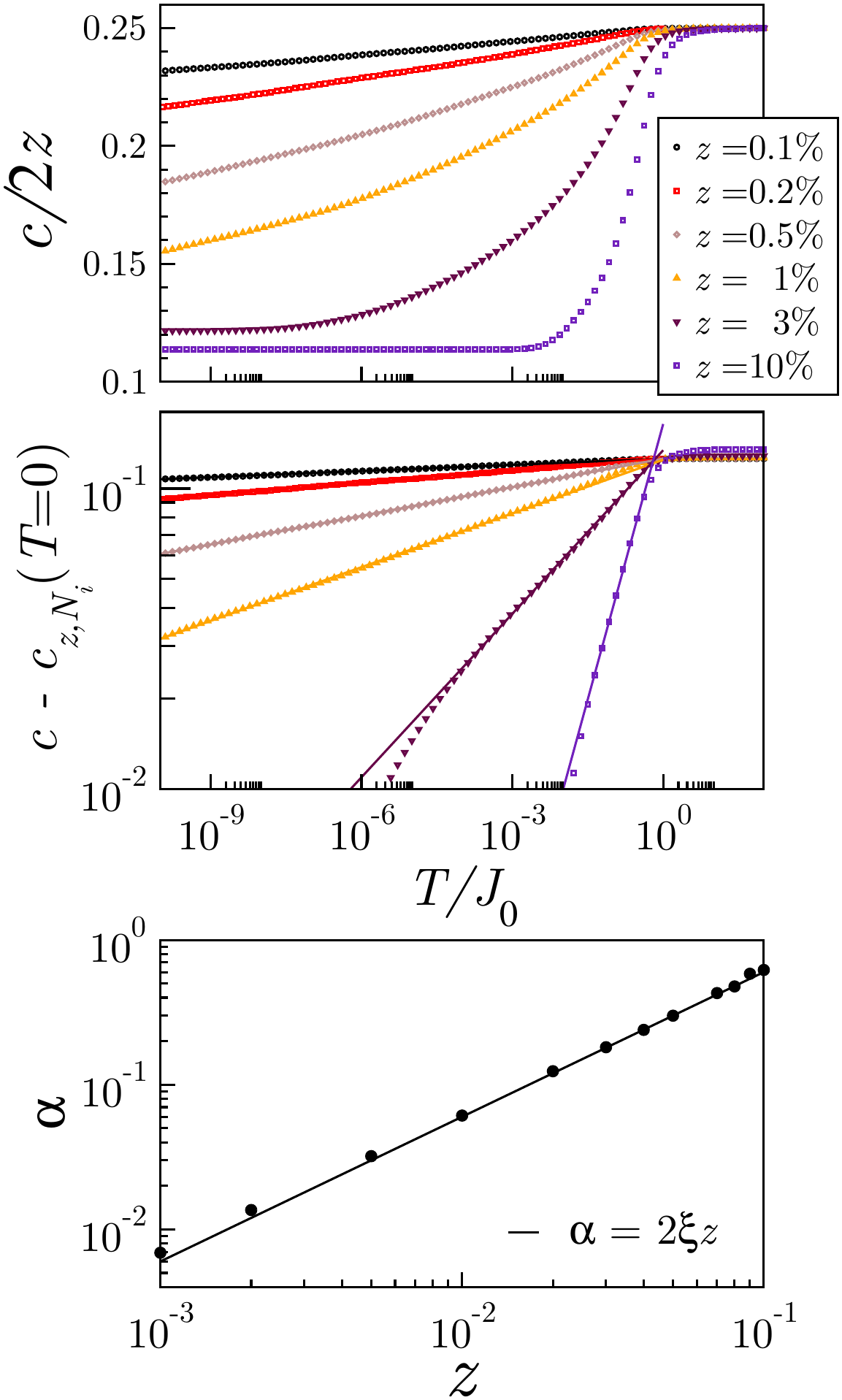}
\caption{(Color online) Exact diagonalization results on the thermal
  behavior of the Curie constant $c(T)$ within the effective model
  description for $N_i=10$ impurities.}
\label{fig:ED-cofT}
\end{figure}

\subsubsection{ED results}

We present on Fig.~\ref{fig:ED-cofT} the results obtained from exact
diagonalization with $N_i=10$ impurities and averaged over 10000
samples. The first remarkable result is that the zero-temperature
plateau is very well approximated on finite sizes by using
\eqref{eq:c-exact-ladder} or its exact numerical evaluation. In fact,
the effective model is not a bipartite lattice model to which the
theorem applies, but the fact that it originates from a bipartite
model to which the theorem applies (without frustration) seems to make
it hold even in the effective model. The reason for that is certainly
that the sign of the couplings satisfy the bipartite nature of the
original lattice. The low-temperature departure from the $T=0$ plateau
is very sensitive to finite-size effects and disorder averaging. This
makes it hard to capture the hypothetical thermodynamical behavior
with this data. Yet, for the intermediate temperatures regime up to
the high-temperature saturation plateau, we obtain a very good fit of
the $c(T)-c(T=0)$ data using a power-law, as one can see from
Fig.~\ref{fig:ED-cofT}(b). Collecting the fitted exponents
Fig.~\ref{fig:ED-cofT}(c) shows a very good agreement with the
$2\doping\xi$ prediction of the random dimer model.

\subsubsection{The RSRG scenario from the F-AF random chain}

However, the behavior in the thermodynamical limit within the
effective model is difficult to address. To sketch a possible
scenario, we refer to the works done on the F-AF random
chain~\cite{Furusaki1995, Westerberg1997, Frischmuth1997,
  Frischmuth1999}. Indeed, for reasonably short correlation lengths
$\xi$, the effective model should fall into the F-AF universality
class in the RSRG sense. This universality class has been dubbed as
the large-spin phase, which is of the Griffith's type, and for which
it has been found that the total spin follows the random-walk scenario
discussed above, and that a power-law correction to the
zero-temperature Curie constant is expected. As regards the possible
universal exponents of this phase, it was found
numerically~\cite{Westerberg1997} that it is strongly dependent on the
singular nature of the initial coupling distribution. By denoting
$P(J) \sim |J|^{-y}$ the initial distribution of the couplings, the
following scenario is proposed: (i) when $y>y_c$ with $y_c \simeq
0.7$, the RSRG flows towards a non-universal fixed point with
non-universal value of $\alpha$. This exponent should depend on
$\doping$ and $\xi$ but is not necessarily equal to $2\doping\xi$;
(ii) when $y<y_c$ (initial distribution ``not too singular''), the
RSRG flows towards a universal fixed point with $\alpha \simeq 0.22$.
QMC calculations~\cite{Frischmuth1997, Frischmuth1999} have
demonstrated the following typical behavior for the Curie constant on
the F-AF random chain: at high-temperatures below the saturation
plateau, $c(T)$ strongly depends on the initial distribution coupling.
Yet, at low enough temperatures, the various $c(T)$ curves collapse
very close to the RSRG prediction with $c(T)-c(0) =
K(T/J_{\text{max}})^{0.21}$ where $c(0)=1/12$ and $K \simeq 0.117$.

Coming back to the situation of doped ladders, we may propose the
following scenario. Provided the RSRG picture is applicable to the
ladder, something certainly true for $\doping \xi \ll 1$ but hard to
justify when $\doping \xi \sim 1$, we first expect from
Refs.~\onlinecite{Frischmuth1997, Frischmuth1999} that the
high-temperature regime is always dependent on the
distribution. Interestingly, in the doped ladder situation and within
the random dimer picture, we found that this regime displays a
power-law behavior with an exponent $2\doping\xi$ which is simply
related to the coupling distribution exponent. Then, one expects that
the RSRG picture develops at low-temperatures with two possibles
cases.  The $y_c \simeq 0.7$ criteria translates on ladder to a
critical doping $\doping_c \simeq 1-e^{-0.15/\xi}$ such that (i) if
$\doping < \doping_c$, the low-temperature exponent is non-universal,
dependent on $\doping$ and $\xi$ and could differ from the
high-temperature exponent expected to be $2\doping\xi$; (ii) if
$\doping > \doping_c$, one can fall into the RSRG universality class
and the low-temperature exponent should become independent of
$\doping$ and $\xi$ and reaches $0.22$. One must notice that the
second situation can actually be realistic for doped ladders in the
weakly coupled regime, since, for instance, $\xi \gtrsim 7$ for
$J_{\perp} = J_{\parallel}/2$, giving $z_c \simeq 2\%$.

\begin{figure}[t]
\centering
\includegraphics[width=\columnwidth,clip]{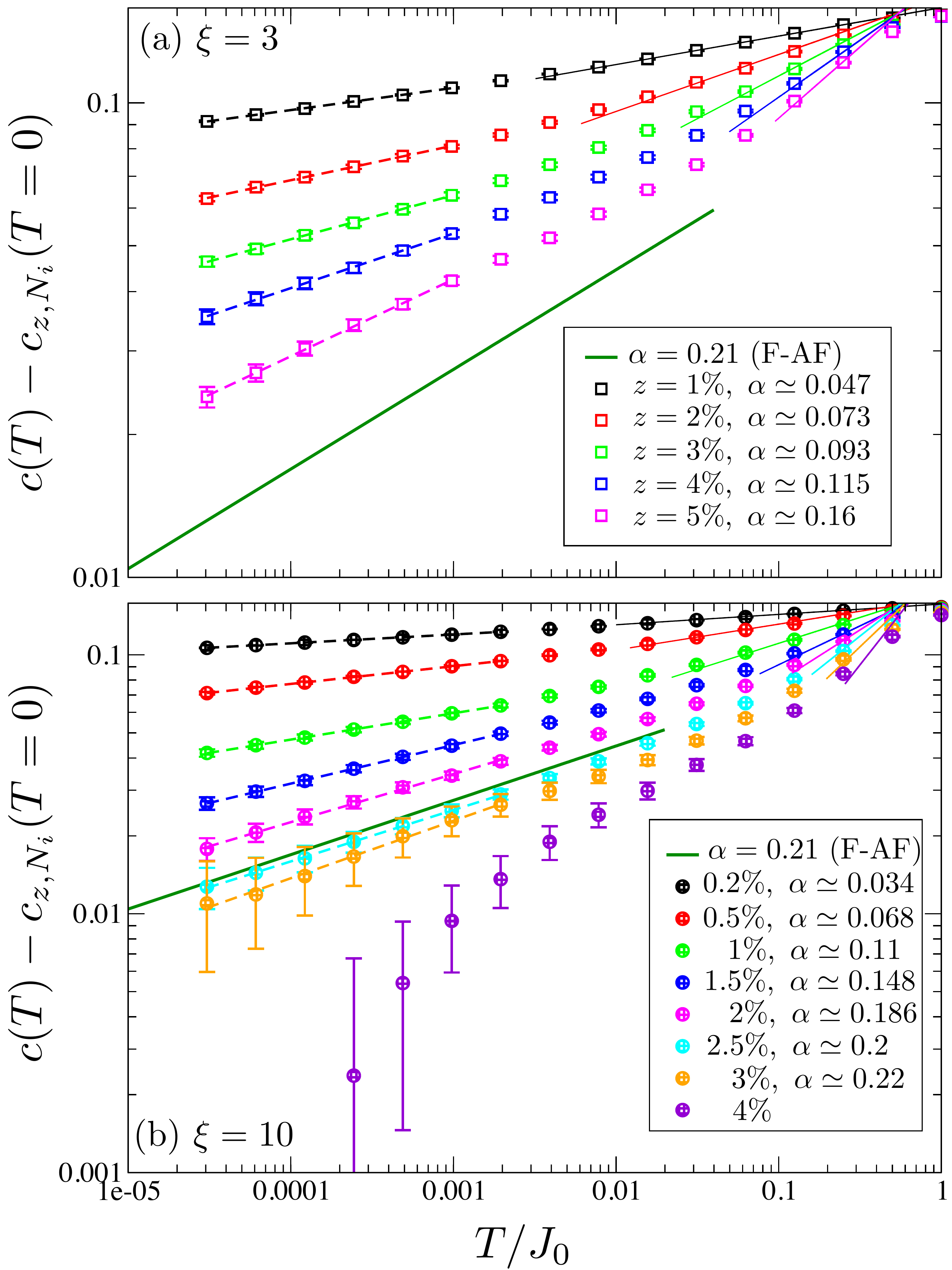}
\caption{(Color online) QMC results on the effective model with
  $N_i=100$ impurities and two correlation length $\xi=3$ (a) and
  $\xi=10$ (b). Disorder averaging has been done over a few thousands
  of random configurations. The Curie constant per impurity (from
  which the asymptotic value $C_{z,N_i}/N_i$ has been subtracted) is
  shown {\it{vs.}} temperature for various concentrations
  (symbols), together with power-law fits (dashed lines) of the form
  $T^{\alpha}$ where $\alpha(z)$ is a varying exponent indicated on
  the plot. The lines in the high-temperature region are with exponent
  $2\xi\doping$. The $\alpha=0.21$ line is the universal regime found
  in the F-AF chain by Frischmuth
  \emph{et~al.}~\cite{Frischmuth1997,Frischmuth1999}, including the
  same prefactor $K\simeq 0.117$.}
\label{fig:QMC-effective-model}
\end{figure}

\begin{figure*}[t]
\centering
\includegraphics[width=\textwidth,clip]{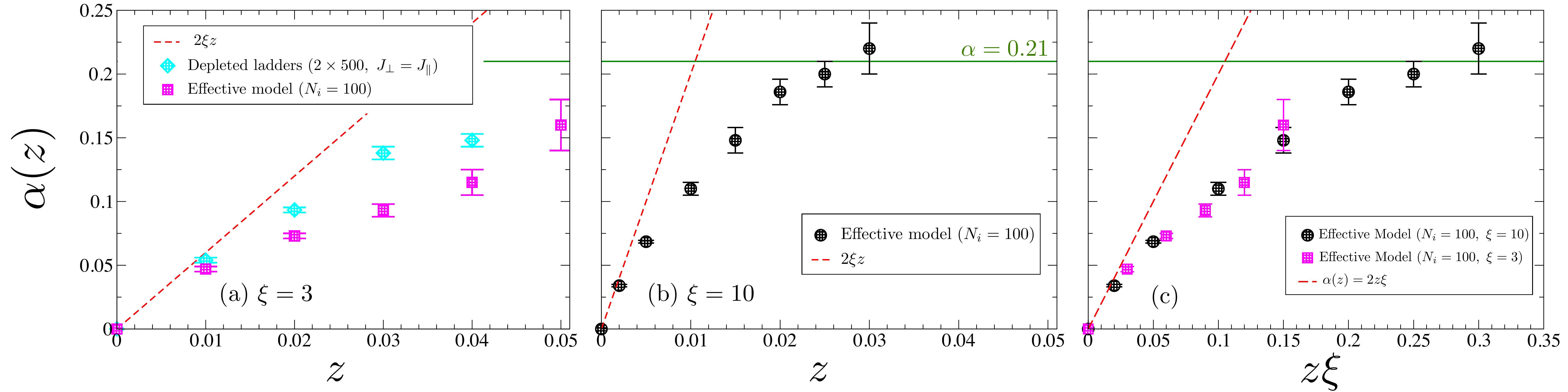}
\caption{(Color online) QMC estimates for the Curie constant exponent
  $\alpha(z)$ from the effective model with $N_i=100$ impurities with
  (a) $\xi=3$ and (b) $\xi=10$. The horizontal $\alpha=0.21$ lines
  indicate the RSRG universal regime of the F-AF chain. In panel (a),
  results for the depleted isotropic ladder are shown for
  comparison. The two cases $\xi=3,~10$ are also shown in panel (c)
  {\it{vs.}} $z\xi$ where results seem to collapse on a single
  curve. In all cases, the exponent $\alpha(z)$ is smaller than the
  prediction $2z\xi$ which seems to hold only in the limit $z\xi\to
  0$.}
\label{fig:QMC-effective-model-2}
\end{figure*}

\subsubsection{QMC results}

In order to test this scenario which relies on several questionable
assumptions, although it looks plausible within the usual pictures
discussed in random one-dimensional magnets, we have carried out QMC
simulations on the effective model up to $N_i=100$ impurities.  The
results for the Curie constant are plotted in
Fig.~\ref{fig:QMC-effective-model}(a-b) for two values of the
correlation length $\xi=3$ and $\xi=10$ which respectively correspond
to $z_c \simeq 0.048$ and $z_c \simeq 0.015$. We observe on these data
a crossover from a fast decaying high-temperature regime, roughly
controlled by the exponent $2\doping\xi$ and a smaller
doping-dependent exponent at lower temperatures. For large values of
$\doping\xi$, the deviation is even clearer and the exponent does not
seem to exceed the RSRG universal result of $0.21$. We also show the
F-AF universal result on the same plot showing that data at large
$\xi\doping$ qualitatively saturates on this limit. These results give
good confidence that the above scenario is plausible.

To further test the scenario, we have extracted the low-temperature
exponent and plot it against $\doping$ and $\xi\doping$ on
Fig.~\ref{fig:QMC-effective-model-2}. We observe that the low-doping
regime is consistent with the $2\xi\doping$ limit while intermediate
dopings display significant deviations and a tendency to saturate
around the RSRG universal regime for $\doping \gtrsim \doping_c$.
Yet, the validity of the random F-AF RSRG picture can be questioned
for two main reasons: when $\doping\xi$ becomes large, the dilute
short range interaction limit fails and it is not guaranteed that the
RSRG is still under control with longer range interaction. Second, as
we will see on the magnetic curve, the discretized nature of the
distribution can play a quantitative role. The criteria for the
initial distribution exponent $y$ is valid within a continuous
description but the discretized nature of the coupling can make the
distribution more singular. Interpretation in that sense was proposed
on the same model through the study on correlation
lengths\cite{Trinh2012}. Still, we observe that the RSRG argument does
capture a lowering of the $\alpha(\doping)$ curve w.r.t. the
$2\xi\doping$ naive expectation. Having an accurate quantitative
description of this curve yet remains a challenging question.

\subsubsection{Effect of frustration and incommensurability}

Frustration makes the lattice non-bipartite so that the exact results
\eqref{eq:c-exact-ladder} do not apply. Still, within the effective
model, if frustration is not too strong, the system remains
commensurate and the above results remain valid and shows that the
behavior is the same. When frustration is large enough to induce
incommensurability in the system, the effective model is affected and
next-nearest neighbor couplings can become frustrating. In this
situation, QMC calculations are not possible due to the sign problem
and we carry out ED calculations, limited to a $N_i=10$ impurities. We
do not show the data because the picture remains essentially and
quantitatively the same as for the commensurate regime. This absence
of strong qualitative differences is certainly due to the fact that
the spin correlation length is small and prevents frustrating effects
to develop on large scales. Furthermore, RSRG arguments tell that the
commensurate and incommensurate cases should fall into the same F-AF
random chain picture so that incommensurability does not actually
plays a fundamental role in this model, at least on its
one-dimensional version.

\begin{figure*}[t]
\centering
\includegraphics[width=\textwidth,clip]{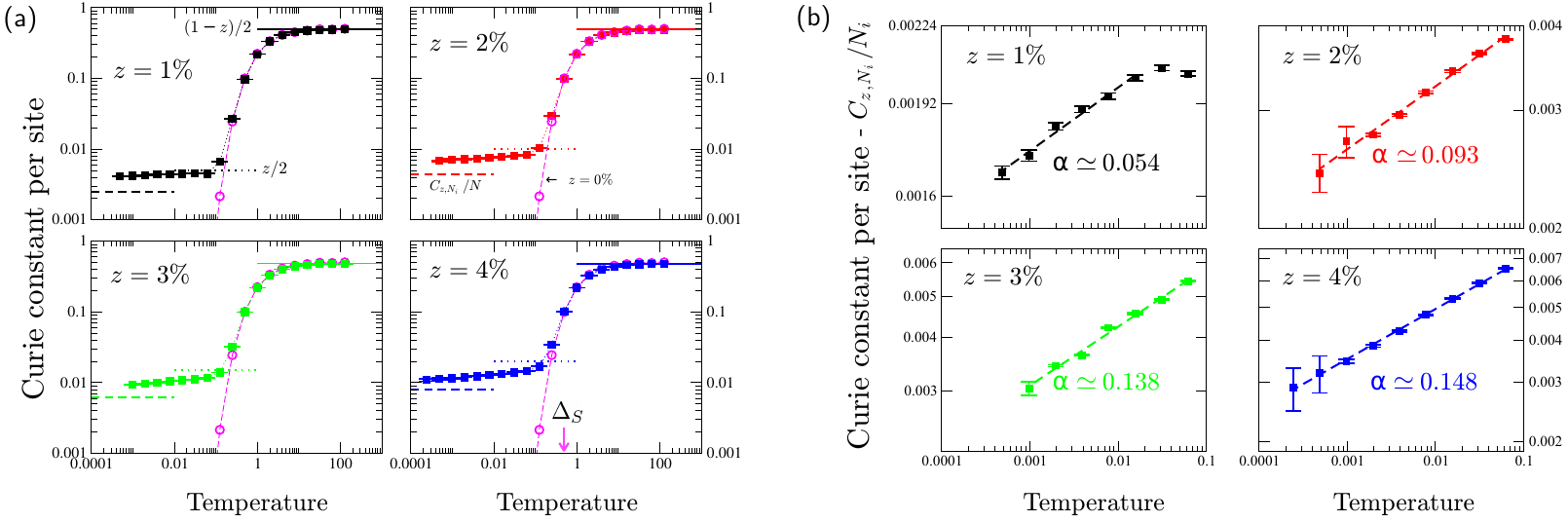}
\caption{(Color online) Finite temperature QMC results for the Curie
  constant per site of randomly depleted 2-leg $S=1/2$ isotropic
  ladders ($J_\perp=J_s=1$) of total size $N=2\times 500$ sites with
  $N_i=10,~20,~30,~40$ non-magnetic impurities, corresponding to
  $z=1,~2,~3,~4\%$, as indicated on the plots. QMC data have been
  averaged over several thousands of independent disordered
  samples. (a) The Curie constants are shown for the entire
  temperature regime. Full lines show the high temperature free spin
  regime $c(T)=(1-z)/4$; dotted lines the intermediate plateau regime
  at $z/4$ (roughly visible for $z=1\%$ but absent for higher
  dopings); and dashed lines the expected very low-T limit
  $C_{z,N_i}/N_i$ computed in section~\ref{sec:CzNi}. The open magenta
  symbols show the Curie constant per site of the clean case which
  start to deviate, rapidly falling to zero, below the spin gap
  $\Delta$ (arrow on the $z=4\%$ subpanel. (b) same data as in the
  left part from which the asymptotic Curie constant per site
  $C_{z,N_i}/N_i$ has been subtracted. One sees that the low-T part
  slowly goes to zero as power-laws, with exponents $\alpha(z)$
  indicated on the plot.}
\label{fig:QMC-cofT-full}
\end{figure*}

\subsection{The full $c(T)$ curve from the microscopic model}

We now discuss the overall behavior of $c(T)$ computed on the original
microscopic model doped with impurities in order to test the scenario
by Sigrist and Furusaki discussed above. We take the situation of an
isotropic ladder which has $\xi \simeq 3$ but in which the energy
scales $J_0$, $J_{\text{max}}$ and $J_1$ are very close to each other
(see Fig.~\ref{fig:J0}). The first significant effect as seen on
Fig.~\ref{fig:QMC-cofT-full}(a) is thus the absence of an intermediate
plateau of independent impurity spins. Thermal magnons are activated
before impurity spins become uncorrelated by thermal excitations so
that the contributions of both can never be separated. This absence of
plateau in the isotropic ladder will have its counterpart in the
magnetic curve while scanning the energy scales with the magnetic
field rather than with the thermal energy (see
Sec.~\ref{sec:magnetization}). In the large $J_{\perp}$ limit, the
separation of energy scales suggests that the plateau could be visible
but we have not checked it numerically.

At temperatures slightly below the temperature corresponding to the
spin gap $T=\Delta_s$, a power-law behavior is clearly visible showing
the regime in which the effective model accounts for the physics. The
exponent is found to depend on doping, with very small exponents at
low dopings which could give the impression of the presence of a
plateau, although this is not correct. A systematic extraction of the
exponent (see Fig.~\ref{fig:QMC-cofT-full}(b)) gives the results
plotted on Fig.~\ref{fig:QMC-effective-model-2} against the effective
model results. Slightly larger exponents are found but the agreement
can be viewed as correct considering the low values of the exponents
and the difficulty to tackle this low-temperature regime
numerically. Consequently, the effective model seems to capture the
physics at low-energy of the interacting impurity spins. Although the
convergence towards the universal RSRG regime is plausible at small
$J_{\perp}$ and low doping from our results on the effective model,
the numerical challenge it represents on the microscopic model is
beyond the scope of this paper (the spin gap becomes significantly
smaller).

\section{Magnetic curve and deviations from Brillouin's behavior}
\label{sec:magnetization}

Another way to probe the effective interactions between impurities is
to scan the energies using a magnetic field rather than
temperature. As we are studying the part of the magnetic curve which
typically lies below the spin gap $\Delta_s$, the results correspond
to accessible magnetic fields, and are then particularly relevant to
experimental measurements. We aim at proposing some possible relevant
fits of this regime. Above the spin gap, the elementary excitations
involve magnons which can localize in the disordered
environment. There, the physics becomes quite different and we do not
address these questions related to Bose-glass physics.

\subsection{The zero-temperature magnetization jump and saturation plateau}

We now turn to the generic behavior of the magnetic curve $m(H,T)$. In
the previous section, the non-trivial behavior when $H\rightarrow 0$
at finite $T$ was discussed. Physically, it corresponded to
susceptibility measurements performed with $H \ll T$. Strictly
speaking, we must have $m(H,T)=0$ when $H=0$ due to the SU(2)
symmetry. If one now considers a finite-system with $T=0$ and a small
but finite magnetic field, the degeneracy within a sector of total
spin $S$ will be lifted to favor the state $S^z=S$. Then, there exists
a disorder averaged magnetization jump $\delta m = m(H=0^+,T=0) -
m(H=0,T=0)$ which matches $\delta m = 2\overline{\moy{S_z}}/N =
2\overline{S}/N$. This is typical of a partially ferromagnetic state.
If one does not take into account chain breaking effects, as we do for
the effective model, the scaling of $\overline{S}$ yields a
magnetization jump that vanishes in the thermodynamical limit as
\begin{equation}
\delta m \simeq \sqrt{\frac{2\doping(1-\doping)}{\pi}}\frac{1}{\sqrt{N}}\;.
\label{eq:deltam}
\end{equation}
Interestingly, we notice that, due to the random walk argument, the
prefactor is actually related to the zero-temperature Curie-constant
$c$ by $\delta m \sim \sqrt{c/N}$ which makes a connection between the
two non-commutating limits of the magnetic responses under study. If
we take chain breaking effects into account, then there exists a jump
\emph{even in the thermodynamical limit} which reads $\delta m
=\overline{\overline{s}} \sim \frac{\sqrt3}2 \doping^{3/2} \sim
c^{3/2}$ in the diluted limit $\doping\ll1$.
 
Lastly, one expects that, within a picture of impurities bringing each
exactly one spin, the saturation plateau corresponding to the
polarization of all these spins equals $m = \doping$. Yet, chain
breaking effects should lower this value since configurations where
two impurities are on the same rung do not bring any free spin. Taking
this effect into account gives an expected saturation at $m =
\doping(1-\doping)$. This effect matters for DMRG or QMC data as well
as experiments.

\subsection{Zero-temperature magnetic curves}

\subsubsection{Hints from the random dimer model}

Using the random dimer model, the low-temperature magnetization curve
(for $H\gg T$) takes a Fermi-Dirac form to a good approximation
\begin{equation}
m(H,T) \simeq \doping \int dJ P(J) \frac{1}{e^{(J-H)/T} + 1} \;,
\label{eq:m-Fermi-Dirac}
\end{equation}
where $H$ plays the role of the chemical potential. This is physically
transparent as the system is equivalent in this limit to a collection
of two-level systems with only the singlet and triplet $S_z=1$ states
contributing to the low-energy physics which naturally maps onto
fermionic statistics.

In particular, the $T=0$ limit of this model gives that the magnetic
curve is simply related to the repartition function $R(J)$ of the
couplings through $m(H,T=0) = \doping R(H)$.  In the case of the
continuous distribution \eqref{eq:PJ-powerlaw}, this yields a
power-law behavior
\begin{equation}
m(H,T=0) = \frac{\doping} {2} \bigg ( 1 + \bigg(\frac{H}{J_0}\bigg)^{\alpha} \bigg)\;,
\label{eq:m-RDM-continuous}
\end{equation}
with $\alpha = 2\doping\xi$ and for $H\leq J_0$, which already
deviates significantly from the Brillouin picture. It is important to
notice that situation where $\doping > \doping^*$ from
\eqref{eq:z-star} is physical in the case of systems with a large
correlation length $\xi \simeq 10$. Then, the curvature of the
magnetic curve is expected to change from concave to convex.

\begin{figure}[t]
\centering
\includegraphics[width=\columnwidth,clip]{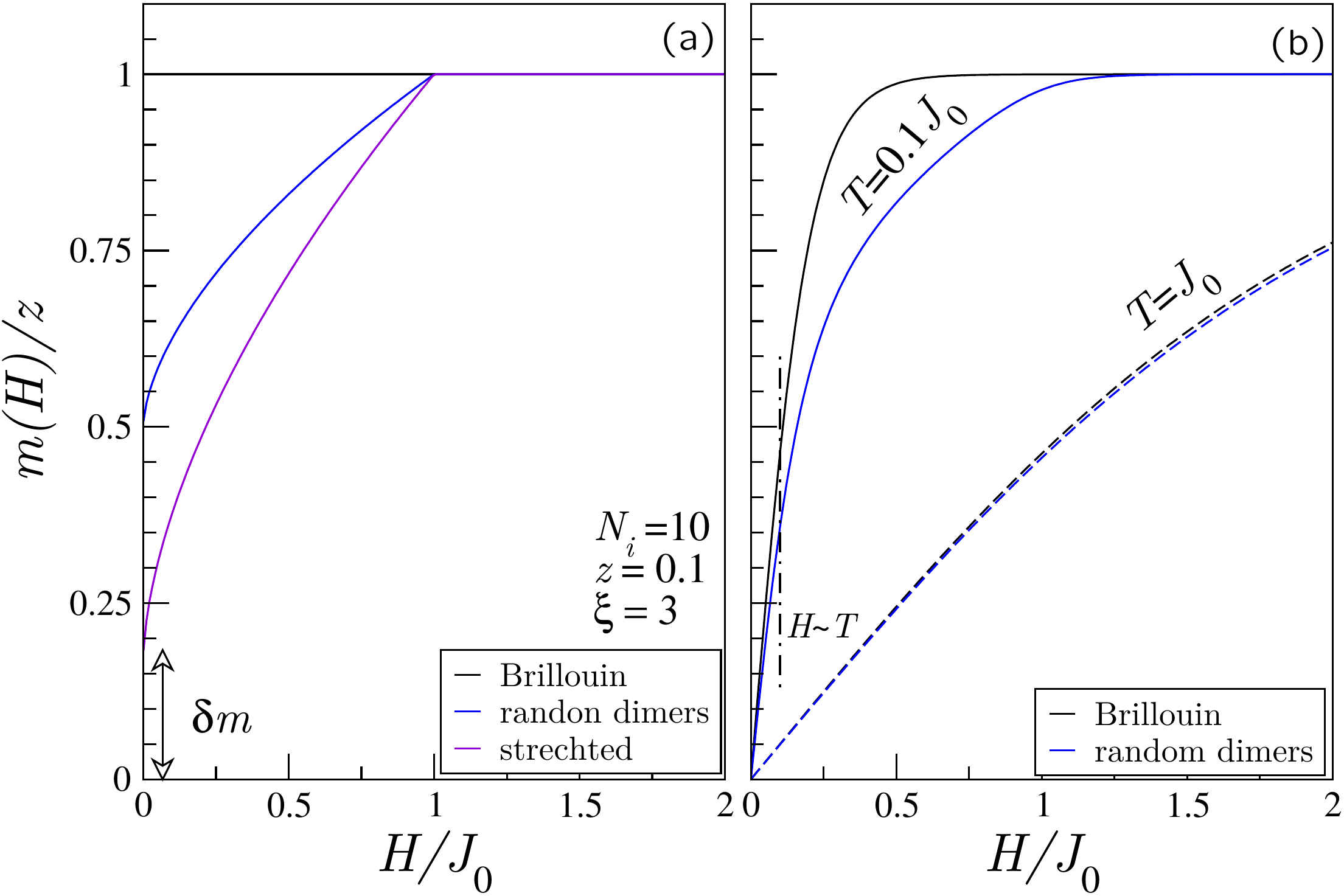}
\caption{(Color online) Schematic behavior of the magnetic curve from
  the random dimer picture at $T=0$ \textsf{(a)} and finite
  temperature \textsf{(b)}. $\delta m$ represents the magnetization
  jump.}
\label{fig:magnetization-RDM}
\end{figure}

Still, we see that the random dimer model fails to reproduce the
correct $H\rightarrow 0$ limit and gives for the jump $\delta m =
R(0)$ ($\delta m = 1/2$ for Eq.~\eqref{eq:m-RDM-continuous}). One can
incorporate the exact result \eqref{eq:deltam} in the RDM by
stretching the repartition function of antiferromagnetic couplings
$R_+(J)$. We thus define the phenomenological ''stretched random
dimer'' ansatz as
\begin{equation}
m(H,T=0) = \delta m + (\doping-\delta m) R_+(H)\;.
\label{eq:m-RDM-pheno}
\end{equation}
Physically, the issue of the random dimer model is that it works with
total spins $S=0$ and $S=1$ and cannot capture the large-spin
formation.  These features and ansatz of the random dimer model are
represented on Fig.~\ref{fig:magnetization-RDM}(a). Lastly, this rough
understanding of the shape of the curve leads to the following simple
power-law fit which could be useful for experiments or numerical
calculations:
\begin{equation}
m(H,T=0) = \delta m + (\doping-\delta m)\left(\frac{H}{J_0}\right)^{\alpha}\;,
\label{eq:m-power-law-fit}
\end{equation}
in which one can leave free the three parameters $\delta m$, $J_0$ and
$\alpha$. Interestingly, RSRG arguments~\cite{Westerberg1997} have
also proposed a power-law behavior for the magnetic curve when $H\gg
T$ based on energy scales phenomenology.

\subsubsection{Comparison between effective and microscopic models}

\begin{figure}[t]
\centering
\includegraphics[width=0.9\columnwidth,clip]{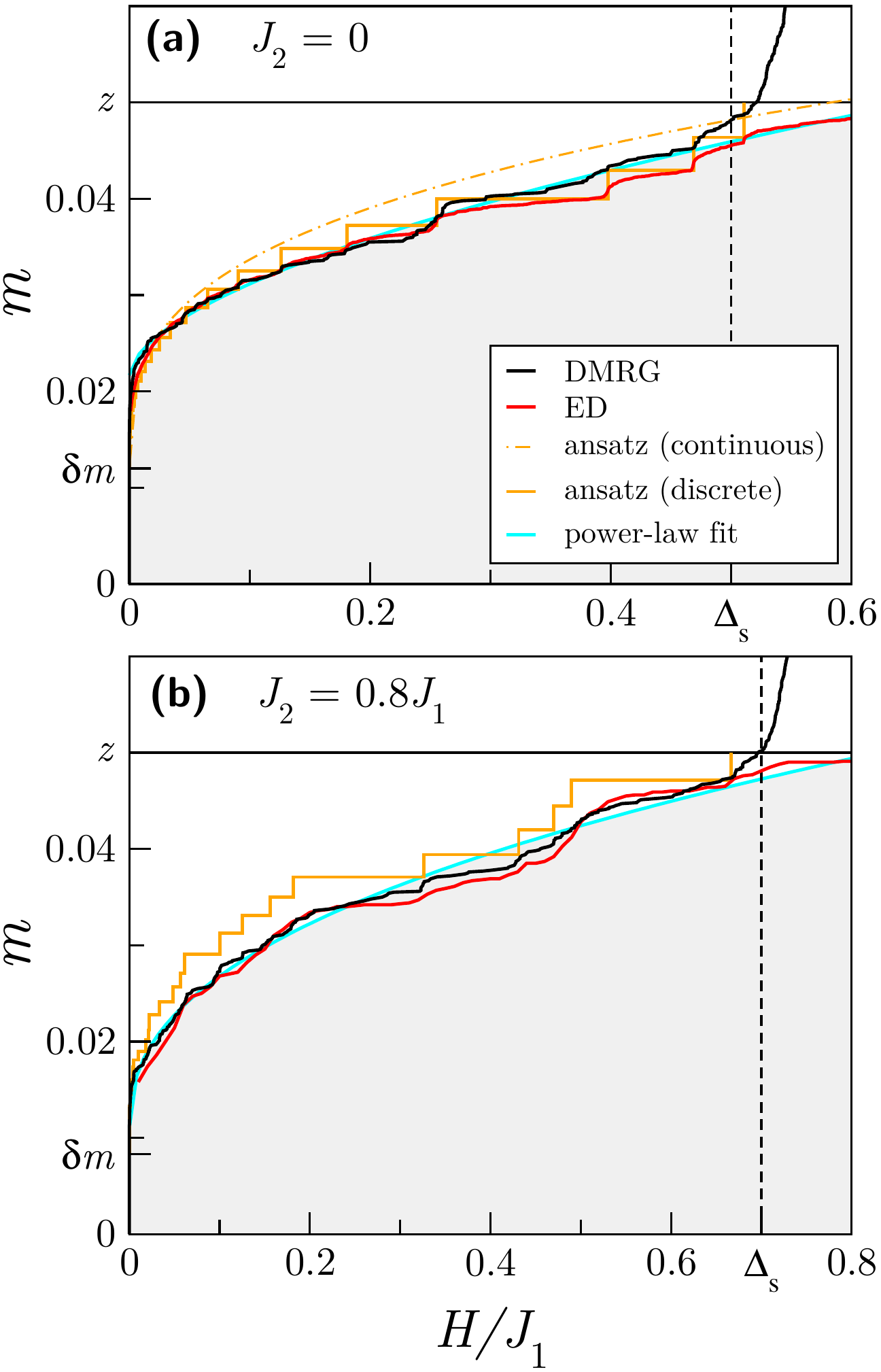}
\caption{(Color online) comparison between DMRG on the microscopic
  model (isotropic ladder) and ED on the effective model. (a) in the
  commensurate regime (b) in the incommensurate regime with
  frustration. The phenomenological ``stretched'' dimer model of
  \eqref{eq:m-RDM-pheno} is also given using either the exact discrete
  or a continuous version of the couplings distribution. The power-law
  fit is done using \eqref{eq:m-power-law-fit} and provides a simple
  account of the deviation from the Brillouin response which is
  emphasized by the grey area. }
\label{fig:magnetization-T0}
\end{figure}

We compute numerically the magnetization curves at zero temperature
using DMRG on the microscopic model and average over many
configurations. The results for the isotropic ladder $J_{\perp}=J_1$
are displayed on Fig.~\ref{fig:magnetization-T0}(a-b) for both a
system without frustration and with frustration in the incommensurate
regime. Qualitatively, the two curve are essentially governed by the
coupling distribution and frustration does not have a drastic
qualitative effect. Interestingly, the simple approximations described
in the preceding section account rather well of the behavior of the
curve. First, the ED on the effective model captures the power-law
like behavior and even underlines the discrete nature of the coupling
distribution. This discrete nature is transparent from the ansatz
\eqref{eq:m-RDM-pheno} using the exact effective couplings. The DMRG
does show faded steps corresponding to the larger couplings, and ED
too. The envelope of the random dimer model is captured by the
continuous version of the coupling distribution. Yet, we see that one
really needs the discretized version to be quantitative. Last, we show
that a fit of the form \eqref{eq:m-power-law-fit} captures the mean
power-law behavior of the curve in a satisfactory way. This is all the
more relevant as we will see that temperature tends to fade the steps
due to the discrete couplings.

One can notice the slight difference between the ED and DMRG
results. We attribute these to two main possible effects. First, as
the systems are chosen to have the same total number of impurities,
the limitation of the two-body interaction effective model can play a
role. Many-impurity interactions could become relevant even though
these are subdominant effects. Second, we have seen that chain
breaking effects must make the saturation plateau occur at
$\doping(1-\doping)$, but it also has the effect of averaging magnetic
curves over various dopings. Indeed, in the presence of chain breaking
effects, each piece has a different doping which approaches $\doping$
on average but can be lower or higher. This should have significant
effects compare to the fixed $\doping$ curve of the ED on the
effective model.

The last important remark is that no saturation plateau is reached in
the isotropic ladder. As for the Curie constant plateau, this is due
to the fact that the typical energy scales are of the same order of
magnitude $J_0 \sim J_{\text{max}} \sim \Delta_s$ (see
Fig.~\ref{fig:J0}). Then, magnons become activated by the magnetic
field before all impurities are truly polarized. The energy scales
separation in the strong-coupling limit suggests that such a plateau
could be possible at large $J_{\perp}$ but we have not investigated
this situation in details. In particular, the microscopic model
displays small couplings in this limit which are harder to capture
with DMRG.

\begin{figure*}[t]
\centering
\includegraphics[width=\textwidth,clip]{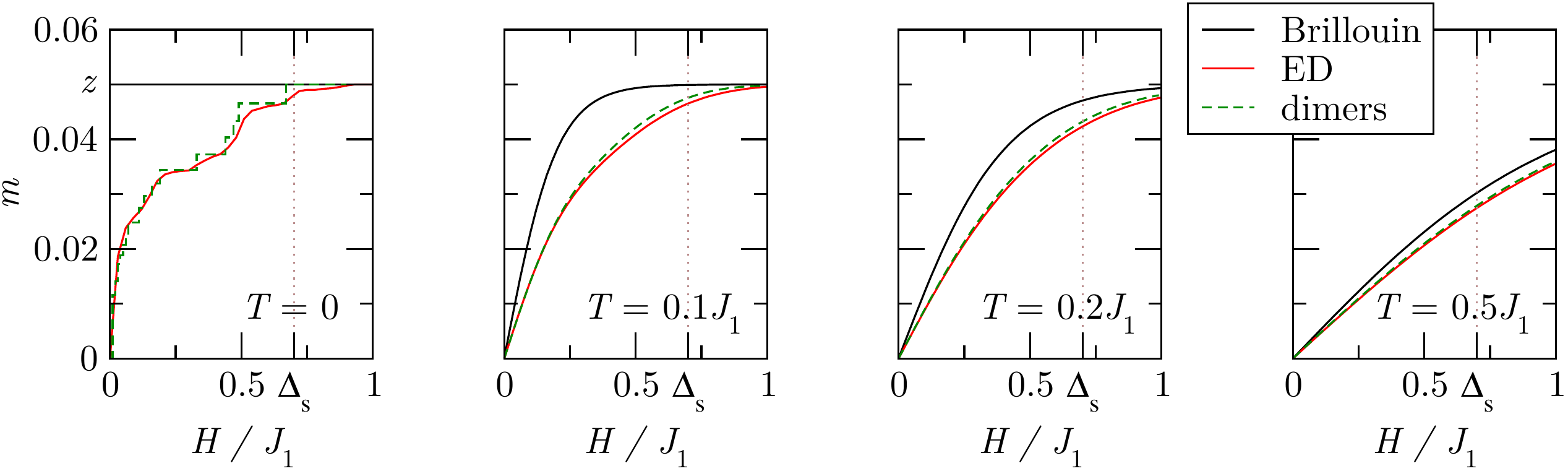}
\caption{(Color online) comparison between the Brillouin result for
  independent spin, the ED curve on the effective model and the
  phenomenological ``stretched'' random dimer model at finite
  temperature from Eq.~ \eqref{eq:m-RDM-pheno-finiteT}. Four
  increasing temperatures are given.}
\label{fig:magnetization-finiteT}
\end{figure*}

\subsection{Finite temperature}

\subsubsection{Random dimer model}

The approximation of Eq.~\eqref{eq:m-Fermi-Dirac}, naturally yields
low-temperature corrections from a Sommerfeld expansion, valid
provided $T \ll H$, which reads $ m(H,T) = m(H,0) + \frac{\pi^2}{6}
P'(H) T^2 + \cdots $, where $P'(H)$ is the derivative of the coupling
distribution. For instance, in the case of the continuous distribution
\eqref{eq:PJ-powerlaw} and taking the approximation $\doping^* \simeq
1/(2\xi)$, the temperature corrections depend on doping and magnetic
field through
\begin{equation*}
m(H,T) = m(H,0) + \frac{\pi^2}{3}\frac{\doping(\doping-\doping^*)}{{\doping^*}^2}\left(\frac{H}{J_0}\right)^{\frac{\doping}{\doping^*}}\left(\frac{T}{H}\right)^{2} \;.
\label{eq:m-sommerfeld-2}
\end{equation*}
Here again, the response to a small temperature is expected to
strongly depend on the side of the limiting case $\doping=\doping^*$,
displaying a change in sign on the corrections.  In the particular
situation where $\doping=\doping^*$, for which $P(J)$ is flat, the
magnetization curve of the random dimer model can actually be computed
exactly
\begin{equation*}
\begin{split}
m(H,T) =& \doping \frac{\sinh[H/T]}{1+2\cosh[H/T]} \\
&\times \frac{T}{J_0}\ln\left\{
\frac{1+e^{J_0/T}(1+2\cosh[H/T])}{1+e^{-J_0/T}(1+2\cosh[H/T])}
\right\}\;. 
\end{split}
\label{eq:m-exact-z1}
\end{equation*}
Following the previous remark on the inability of the random dimer
model to account for the large-spin formation, we can devise an
extension of the stretched dimer model at finite temperatures using
the following ansatz:
\begin{equation}
\begin{split} 
m(H,T)\simeq &\delta m\tanh\left(\frac{H}{2T}\right)\\
&+\frac{z-\delta m}{1-R(0)}\int\limits_{J>0} m_{\text{dimer}}(H,T;J) p(J)dJ\;,
\end{split}    
\label{eq:m-RDM-pheno-finiteT}
\end{equation}
where the first part accounts for the contribution of ferromagnetic
couplings, while the second accounts for the magnetization process of
antiferromagnetic dimers. The first term should in principle
correspond to a Brillouin function of spin $\bar{S}$ but this version
already gives satisfactory results.

\subsubsection{Comparison with ED and QMC}

\begin{figure}[b]
\centering
\includegraphics[width=\columnwidth,clip]{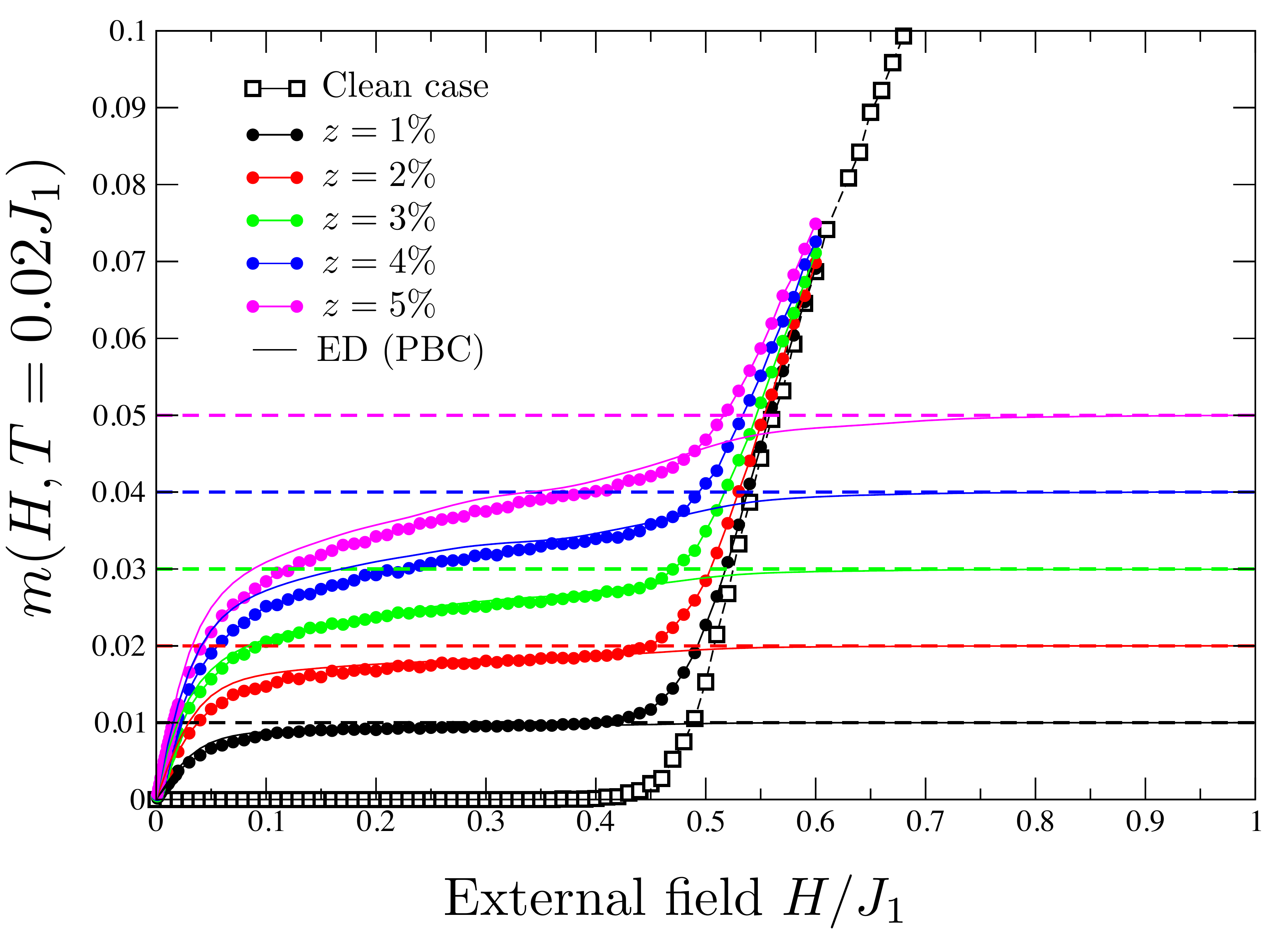}
\caption{(color online) QMC results for the longitudinal magnetization
  {\it vs.} external field $H/J_1$ of depleted ladders of size
  $N=2\times 500$ sites, averaged over $\sim 500$ disordered samples,
  at finite temperature $T/J_1=0.02$. Different impurity
  concentrations $z=1,~2,~3,~4,~5\%$ are shown, together with the
  clean case at the same temperature for comparison. Horizontal dashed
  lines show the expected saturation value for the impurities $m_{\rm
    sat}=z$, and the full lines are ED results obtained with PBC on
  1000 random clusters of 10 impurities from which one clearly sees
  that the saturation value is only reached when $H/J_1\to 1$.}
\label{fig:magnetization}
\end{figure}

The effect of temperature is first discussed on
Fig.~\ref{fig:magnetization-finiteT} by showing the comparison between
the Brillouin response to the ED and random dimer model predictions
for four increasing temperatures. All curves should collapse at high
temperatures $T\gg J_{\text{max}}$. We see that the zero-temperature
steps are rapidly faded as temperature is turn on. Still, the
deviation from the Brillouin curve due to the interaction remains well
visible for finite temperature and actually makes the random dimer
model almost exact.

In order to validate the above comparison, we have compared the ED on
the effective model to QMC, which is the appropriate method for finite
temperature calculations on the microscopic model. In
Fig.~\ref{fig:magnetization}, one observes a rather good agreement for
several realistic dopings. The larger the doping, the larger the
deviation from the Brillouin curve is and the larger the distance from
saturation is when the magnons set in. The slight difference between
ED and QMC could here again be attributed to many-impurities
interactions not taken into account in the effective model and also to
the effective doping averaging induced by chain breaking effects, as
for the zero-temperature curve. Then, the following message is almost
quantitatively correct from the comparison between all different
approaches: the low-part of the magnetic curve probes the couplings
distribution between the impurities. This is evident in the random
dimer model and the picture survives to the microscopic model rather
well. This simple analysis is certainly due to the fact that we are
discussing a simple observable (density of magnetization) which is
little affected by low-energy behavior or correlations in the
system. Therefore, it could be accessible and interesting to test such
phenomenology in experiments working with quasi one-dimensional
systems.

\subsection{Consequences for experiments}

Our theoretical study could in principle apply to several realistic
spin gapped materials. However, as seen above, a clear separation
between different energy scales -- the spin gap $\Delta_s$ below which
free local moments are expected, and the largest effective coupling
$J_{\text{max}}$ below which they start to correlate upon random F-AF
exchanges -- would be difficult to achieve in systems close to the
isotropic ladder limit. The separation remains plausible in the
strong-coupling limit, although we have not investigate this point
quantitatively in this paper. In the isotropic case, a saturation
regime of impurity spins will be hardly detectable. Nevertheless, the
regime of large spin could be detected in Curie tails at low
temperature, provided the three dimensional ordering of induced
moments (expected below temperatures set by three dimensional
couplings) occurs at low enough temperature. In such a respect, a new
analysis of susceptibility data of Zn ($S=0$) or Ni ($S=1$) doped
BiCu$_2$PO$_6$ form Ref.~\onlinecite{Koteswararao2010} may give
interesting results, although the three dimensional ordering of
induced moments occurs below a few Kelvins~\cite{Bobroff09}. Perhaps
more promising is the doped Haldane chain system
Y$_2$BaNiO$_5$~\cite{Tedoldi99,Payen00,Das04} where a very small
inter-chain coupling $\sim 10^{-1}$ K is expected from neutron
scattering~\cite{Xu96} despite a very large spin gap $\Delta_s\simeq
100$ K.

More generally, our study clearly shows that Curie tails, present in
all AF materials even for undoped ones, because of intrinsic defects
or Imry-Ma domain formation with random couplings~\cite{Lavarelo2013},
have to be analyzed perhaps more carefully than what is usually
done. In particular, the assumption of free impurities leading to the
extraction of their concentration $\doping$ through the simple form
$\chi_{\rm imp}=\doping/(4T)$ is not expected to be valid in many
experimental situations.

Regarding the magnetization curve, our work can potentially apply to
many materials where Brillouin-like responses are observed upon
increasing the external field. For spin-gapped systems, the effective
couplings between local moments can strongly renormalize downwards the
Brillouin-like magnetization, and pushes the saturation towards larger
magnetic fields, possibly larger than the spin gap $\Delta_s$. This
means that, at the critical field where magnon excitations start to
appear, not all impurity-induced moments have been saturated. Such a
phenomenology is expected for BiCu$_2$PO$_6$ in a
field~\cite{Tsirlin2010,Casola13}. Nevertheless, for this ladder
material~\cite{Tsirlin2010}, and also for other systems such as the
Herbertsmithite Kagom\'e compound~\cite{Bert07}, the presence of
non-negligible Dzyaloshinskii-Moriya (DM) anisotropies make the
situation much more difficult to analyze since DM terms induce a
finite magnetic response also below the spin gap. The modification of
the Brillouin-like response due to the competition between impurity
physics and DM interactions in an external field at finite temperature
is a very interesting subject, relevant for many realistic systems,
that we leave for future studies.

\section{Conclusion}

The physics of randomly depleted ladder, studied initially in the
seminal work of Sigrist and Furusaki~\cite{Sigrist1996}, offers a
remarkable playground for studying the effect of impurity disorder in
gapped systems without and with frustration. In this contribution to
the field, we improved on several intuitive results of
Ref.~\onlinecite{Sigrist1996} to provide quantitative predictions and
comparison to numerics and adressed the shape of the magnetization
curve.  Based on a detailed analysis of the effective couplings
between impurities and of the corresponding coupling distribution, we
focussed the main two magnetic responses: the zero-field
susceptibility, through the temperature-dependent Curie constant, and
the magnetization curve. The first one is shown to have a non-trivial
power-law behavior at very-low temperature in qualitative agreement
with a RSRG scenario. The high-temperature deviation from free
impurity spins is well captured by a simple random dimer model. This
model also accounts qualitatively well for the magnetization curve for
which we give several phenomenological fits at zero and
finite-temperature which are in good agreement with accurate numerical
calculations. One of the key outcome of this study is that
incommensurability (induced by frustration) plays little role in the
local quantities we looked at. Indeed, the main consequence of
incommensurability is a mere reduction of the zero-temperature
spontaneous magnetization and of the low-temperature limit of the
Curie constant. The situation might be different in higher dimensional
system but the one-dimensional version seems to be in the same
universality class, as expected from RSRG arguments. These predictions
on the magnetic responses could motivate experiments in that direction
since the required temperatures and magnetic fields are accessible for
several compounds.

\subsection*{Acknowledgements}

AL thanks C\'ecile Delaporte and Antoine Channarond for insightful
discussions on Markov chains calculations. AL and GR acknowledge
support from the French ANR program ANR-2011-BS04-012-01 QuDec. NL
acknowledge GENCI (grants 2012-x2012050225 and 2013-x2013050225) and
CALMIP for numerical resources and the French ANR program
ANR-11-IS04-005-01.

\appendix

\section{Correlations and susceptibility in the Bond-order mean-field approximation}
\label{app:BOMF}

In this appendix, we detail the calculation of the spin correlation
functions within the bond-order mean-field theory developed for the
frustrated ladder in Ref.~\onlinecite{Lavarelo2011}. We will use the
same notations as in this reference. The dynamical and static
structure factors of the model has also been addressed recently in
Ref.~\onlinecite{Sugimoto2013}.

\subsection{Notations and useful relations from BOMF}

Within BOMF theory in which the singlet operators on rungs are assumed
to condense $\moy{s_i} = \bar s$, the mean-field Hamiltonian is solved
through a Bogoliubov transformation on the triplet operators in $k$-space
$t_{k,\sigma}$
\begin{equation}
\label{eq:Bogoliubov}
 b_{k\sigma}=u_kt_{k\sigma}+v_kt_{-k\sigma}^\dag\;,
\end{equation}
in which $u_k$ and $v_k$ satisfy $u_k^2-v_k^2=1$. This leads to the
diagonal version of the Hamiltonian
\begin{equation}
\Ham_m=E_0+\sum_k\omega_k b_{k\sigma}^\dag b_{k\sigma}\;,
\end{equation}
where $\omega_k$ is the dispersion relation which depends on $\bar s$
and the chemical potential $\mu$ used to enforce the hard-core nature
of the triplets on rungs. These two parameters are usually computed
self-consistently with numerical methods.
 
In this paper, in order to have tractable analytical formulas, we use
the following approximations, which turn out to be good in the
strong-coupling limit, $\bar s \simeq 1$ and 
\begin{equation}
\omega_k \simeq J_\perp \sqrt{1+\frac{J_1}{J_{\perp}}\cos k + \frac{J_2}{J_{\perp}} \cos2k }\;.
\label{eq:dispersion}
\end{equation}
The zeros of $\omega_k$ extended to the complex plane will control the
singularities of most physical quantities. To this end, we introduce
the polynomial
\begin{equation}
\label{eq:polynome}
P(X) = 2 \frac{J_2}{J_{\perp}} X^2 + \frac{J_1}{J_{\perp}} X + 1-\frac{J_2}{J_{\perp}}\;,
\end{equation}
such that 
\begin{equation}
\omega_k \simeq J_\perp \sqrt{P(\cos k)}\;.
\end{equation}

\subsection{Spin structure factor and real-space correlations}

In Ref.~\onlinecite{Lavarelo2011}, we obtained that the spin structure
factor defined by
\begin{equation}
S_k=\sum_{x=1}^L e^{ikx} \mathcal{S}_x\;,
\end{equation}
where $\mathcal{S}_x =
\moy{(\spin{x,1}-\spin{x,2})\cdot(\spin{1,1}-\spin{1,2})}$ are
real-space spin correlations, is given in the BOMF approximation by
\begin{equation}
S_k=\frac{3\bar s^2}{\sqrt{P(\cos k)}}\;.
\end{equation}
We now give more details on the two commensurable and incommensurable
regimes, limited to the strong-coupling regime. We recall that the
transition occurs at $J_{2,c} \simeq J_1/4$.
\begin{itemize}
\item For $J_2 < J_{2,c}$, $P$ has two real roots, lower than $-1$.
Consequently, $S_k$ has branch cuts and four branching points on the
axis $\Re[k]=\pi$, with imaginary parts denoted by $\pm
1/\xi_\text{spin}^\pm$ that define two correlation lengths
$\xi_\text{spin}^\pm$ ($\xi_\text{spin}^+>\xi_\text{spin}^-$), such
that
\begin{equation}
\label{eq:xi_commensurable}
\xi_\text{spin}^\pm \simeq \text{arcosh}^{-1}\left(\frac{J_1\mp\sqrt{J_1^2-4J_2J_\perp}}{4J_2}\right)\;, 
\end{equation}
in the strong-coupling limit.
\item For $J_2=J_{2,c}$, $P$ factorizes exactly and the square root
disappears in the denominator of $S_k$. There is no longer branch
cuts and the branching points merge to give two poles on the axis
$\Re[k]=\pi$, with imaginary part $\pm1/\xi_\text{spin}$, where
\begin{equation}
\label{eq:xi_transition}
\xi_\text{spin}= \text{arcosh}^{-1}\left(\frac{J_1}{4J_2}\right)\;, 
\end{equation}
\item For $J_2>J_{2,c}$, the roots of $P$ have a non-zero imaginary
part. Consequently, the branching points leaves the axis
$\Re[k]=\pi$.  There coordinates can be written as $\pm q\pm
i\xi_\text{spin}^{-1}$ where $q$ is the incommensurate wave-vector
associated to the real-space correlations and $\xi_\text{spin}$ is
the spin correlation length. In the large $J_\perp$ limit, we obtain
\begin{align}
\label{eq:q}
q&\simeq \arccos\left(-\frac{J_1}{2\sqrt{J_2J_\perp}}\right)\;,\\
\label{eq:xi_incommensurable}
\xi_\text{spin}&\simeq \textrm{arcosh}^{-1}\left(\frac12\sqrt{\frac{J_\perp}{J_2}}\right)\;. 
\end{align}
\end{itemize}

Real-space behavior of the correlation function is recovered after a
Fourier transform of the static structure factor:
\begin{equation}
\mathcal{S}_x = \frac{3\bar s^2}{2\pi}\int_0^{2\pi}\frac{e^{ikx}}{\sqrt{P(\cos k)}}dk\;.
\end{equation}
One cannot easily compute this integral using the theorem of residues,
because of the branch cuts, but one can argue that the behavior in $x$
is essentially controlled by $e^{iz_1x}$ et $e^{iz_2x}$ with $z_1$ and
$z_2$ the singularities of $S_k$ in the upper half plane. Furthermore,
due to the presence of the square-root in the denominator, one may
guess the following asymptotic behavior
\begin{equation}
\int_0^{+\infty} \frac{e^{ikx}}{\sqrt{k^2+\xi^{-2}}} dk
\underset{x\gg\xi}{\sim}\frac{e^{-x/\xi}}{\sqrt x}\;.
\end{equation}
Indeed, as for the $J_1$-$J_2$ chain \cite{White1996}, the $1/\sqrt x$
correction yields better fits of the numerical results. We thus have
the following scenario for the correlation functions:
\begin{itemize}
\item For $J_2<J_{2,c}$, in the commensurate regime:
\begin{equation}
 \mathcal{S}_x \sim \frac{(-1)^x}{\sqrt x}\left(Ae^{-x/\xi_\text{spin}^+}-Be^{-x/\xi_\text{spin}^-}\right)\;,
 \end{equation}
 where $\xi_\text{spin}^\pm$ are given by \eqref{eq:xi_commensurable}
 and $A$ and $B$ are two constants that depend on
 $\xi_\text{spin}^\pm$.
\item For the transition point $J_2=J_{2,c}$, we remark that the
  factorization of the denominator makes the decay purely exponential.
  Then, one expects
 \begin{equation}
\mathcal{S}_x \sim C(-1)^x e^{-x/\xi_\text{spin}}\;,
 \end{equation}
 with $\xi_\text{spin}$ given by \eqref{eq:xi_transition} and $C$ a
 constant depending on$\xi_\text{spin}$.
 \item For $J_2>J_{2,c}$, in the incommensurate regime:
 \begin{equation}
\mathcal{S}_x \sim C'\frac{e^{-x/\xi_\text{spin}}}{\sqrt x}\cos(qx+\phi)\;,
 \end{equation}
 where $q$ and $\xi_\text{spin}$ are given by \eqref{eq:q} and
 \eqref{eq:xi_incommensurable}, and $C'$ and $\phi$ are constant
 depending on $q$ and $\xi_\text{spin}$.
\end{itemize}

\subsection{Susceptibility}

In order to compute the magnetic susceptibility at
$\textbf{k}=(k,\pi)$, one applies a magnetic field corresponding to
the wave-vector $\textbf{k}$
\begin{equation}
 \textbf{H}_\textbf{r}=H\cos(\textbf{k}\cdot\textbf{r})\;\textbf{e}_z\;.
\end{equation}
In the BOMF approximation, the Hamiltonian then reads
\begin{equation}
\begin{split}
 \Ham= &E_0+\sum_{k',\sigma} \omega_{k'} b_{k'\sigma}^\dag b_{k'\sigma}\\
 &-\frac{L}{2}H\bar s \left(u_k-v_k\right)\left(b_{k0}+b_{-k0}+b_{k0}^\dag+b_{-k0}^\dag\right)\;.
\end{split}
\end{equation}
The energy correction is obtained from second order perturbation
theory in $H$ as
\begin{equation}
E \simeq E_0-L\bar s^2 \frac{(u_k-v_k)^2}{2\omega_k}H^2\;,
\end{equation}
By definition, the susceptibility $\chi_\textbf{k}$ enters in the
expression through linear response theory
\begin{equation}
 E\simeq E_0-L(\chi_\textbf{k}+\chi_\textbf{-k})H^2\;,
\end{equation}
from which we deduce the following expression for the static
susceptibility:
\begin{equation}
\label{eq:susceptibilite_BOMF}
\chi_{k,\pi} = \frac{\bar s^2}{J_{\perp}-4\mu}\cdot\frac{1}{P(\cos k)}\;,
\end{equation}
where $P$ is the polynom defined in \eqref{eq:polynome}.

\subsubsection{Magnetization profile}

Using the result for the susceptibility, on gets for the magnetization
profile the prediction
\begin{equation}
\moy{S^z_{x,y}}\simeq\frac18(-1)^y\frac{1}{2\pi}\int_0^{2\pi}\frac{e^{ikx}}{P(\cos k)}dk\;,
\end{equation}
where the position $(x,y)$ are relative to the impurity site.  The
integral can be computed with help of the residues theorem applied
over a rectangle of base between $0$ and $2\pi$ and infinite in the
vertical direction.

\section{Couplings distribution in the incommensurate case}
\label{app:PJ-inc}

Performing the change of variables requires the calculation of the
derivative
\begin{align}
\frac{dJ}{dr} & = -\frac{J(r)}{\xi}\left[ 1 + q\xi \tan(qr+\phi)\right] \\
& = -\frac{J(r)}{\xi}\left[ 1 + q\xi \sqrt{\left(\frac{J_0}{J(r)}\right)^2 e^{-2r/\xi} -1 }\right]
\label{eq:derivative}
\end{align}
The zeros of the derivative of denoted by $r_m$ and satisfy the
equation
\begin{equation}
\tan(qr_m+\phi) = -\frac{1}{q\xi} = -\tan \theta
\end{equation}
with $\theta = \arctan{\left(\frac{1}{q\xi}\right)} \in
[0,\pi/2]$. Consequently, we have the zeros
\begin{equation}
r_m = (m\pi - \theta-\phi)/q > 0\;\text{with}\;m=1,2,3,\ldots
\end{equation}
One can take by convention $r_0 = 0$ to define the intervals in which
the sign of the derivative is constant $I_m = [r_m,r_{m+1}]$. It is
clear that the intervals have the same size $r_{m+1} - r_m = \pi/q$.
Formally, one can write
\begin{equation*}
\delta(J-J(r)) = \sum_{r^*_k | J(r^*_k) = J } \left\vert\frac{dJ}{dr}\right\vert_{r=r^*_k}^{-1} \delta(r-r^*_k)\;.
\end{equation*}
The solutions $r^*_k(J)$ of the equation $J(r) = J$ are not
analytically computable in general. For a given $J$, there is at most
one solution in each interval $I_m$ and there is a least one solution
for $|J|\leq J_0$. Let us denote by $S(J)$ the number of solutions at
a given $J$ so that the index ranges $1 \leq k \leq S(J)$.
Using \eqref{eq:derivative}, we have
\begin{equation*}
\delta(J-J(r)) = \sum_{k=1}^{S(J)} \frac{\xi}{|J| + q\xi \sqrt{ J_0^2 e^{-2r^*_k(J)/\xi} - J^2 }}\delta(r-r^*_k(J)) \;.
\end{equation*}

We restrict the discussion to the continuous distribution case since
the analytical formula for the discrete version do not help with
respect to a direct numerical sampling. In this case, the weighting
by the continuous approximation for $p(\mathbf{r})$ gives
\begin{equation*}
P(J) = \doping\xi \sum_{k=1}^{S(J)} \frac{e^{-2\doping r^*_k(J)}}{|J| + q\xi \sqrt{ J_0^2 e^{-2r^*_k(J)/\xi} - J^2 }} 
\end{equation*}
The reduction of probability of large $J$ is understood by
 studying the situation where $J \lesssim J_0$
so that there is only a single solution $r^*_1(J)$. Then, one can write
\begin{equation}
r^*_1(J) = \xi \ln\left(\frac{J_0}{|J|}\right) - \delta r
\end{equation}
with $\delta r > 0$ since the effect of $q$ is to decrease the
position of the solution w.r.t. the $q=\pi$ result. This gives
\begin{equation*}
P(J) = P_{q=\pi}(J) \frac{e^{2\doping\delta r(J)}}{1 + q\xi \sqrt{ e^{2\delta r(J)/\xi} - 1 }}\;.
\end{equation*}
Although it is not obvious in the formula, one may convince one-self
graphically that $P(J) < P_{q=\pi}(J)$ corresponding to a decrease of
the weight at large $J$. Consequently, the weight of small $J$s
increase since the signal can approach zero at any distance.

\section{Distribution of ladder sizes}
\label{app:breaks}

In this section, we study the distribution of sizes of disconnected
ladders $\rho(\ell)$ for a given impurity doping $\doping$.  If we
consider an impurity at position $(x,0)$, there are three positions
for a second impurity that break the ladder : $(x-1,1)$, $(x,1)$ and
$(x+1,1)$. In the diluted limit $\doping\ll1$, the density of cuts is
then $3\doping^2$ and the average length of disconnected ladders
$\bar\ell\simeq1/3\doping^2$. As the cuts are not correlated (at least
at large enough distances), it is reasonable to assume that the
number of cuts follows a geometric law of parameter
$\zeta\simeq3\doping^2$ :
\begin{equation}
\rho(\ell)\simeq\zeta(1-\zeta)^\ell\;.
\end{equation}

\begin{figure}[!htbp]
\centering
\includegraphics[width=0.7\columnwidth,clip]{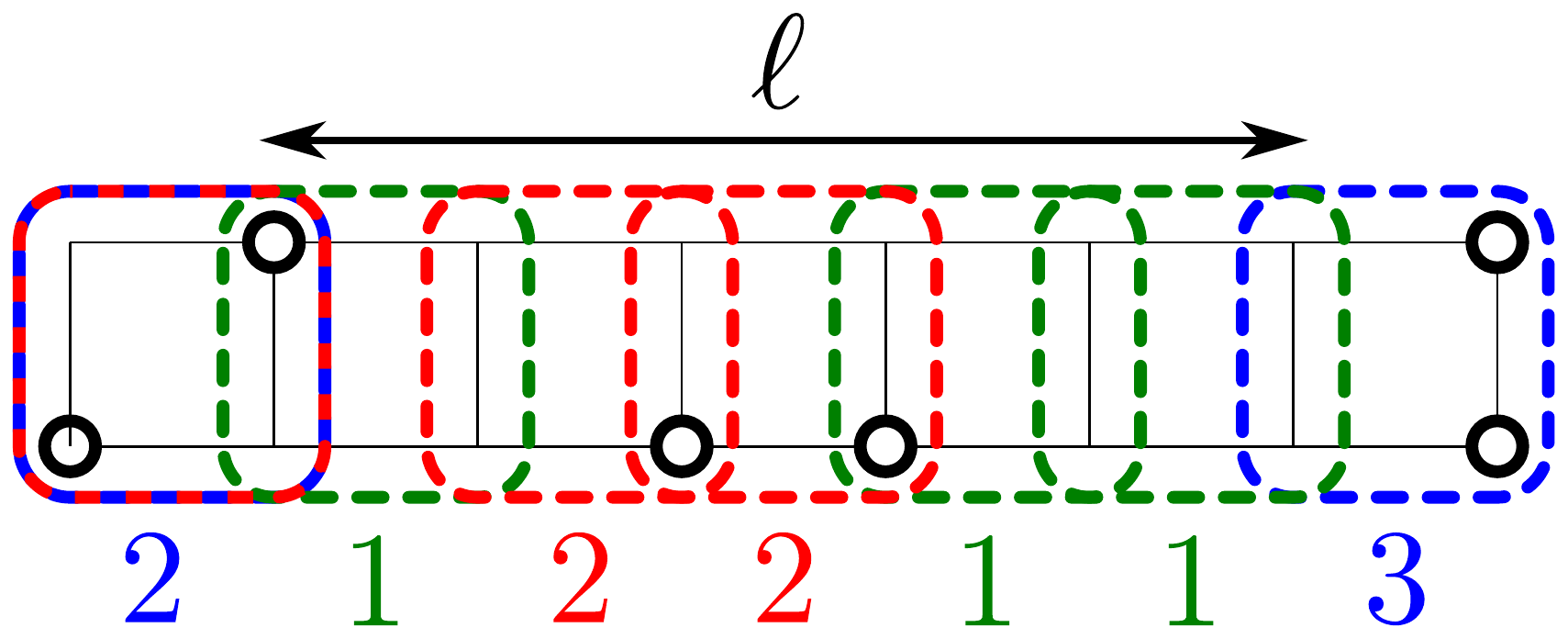}
\label{fig:markov}
\caption{(color online) Schematics of the Markov chain process for chain breaks in the ladder.}
\end{figure}

In fact, the distribution can be calculated exactly. For this, the
ladder is described by a Markov chain $\left(X_n\right)_{n\geq0}$,
where $X_n$ represents the configuration of the plaquette made of the
two consecutive rungs $n$ and $n+1$ (see Fig.~\ref{fig:markov}). 
The Markov property is verified :
the configuration on plaquette $n+1$ only depends of the configuration
on plaquette $n$ as they have a rung in common. The $2^4$
configurations on a plaquette are classified as follows :
\begin{enumerate}
 \item the plaquette dos not break the ladder and there is no impurity on the second rung,
 \item the plaquette dos not break the ladder and there is one impurity on the second rung,
 \item the plaquette breaks the ladder.
\end{enumerate}
The transition matrix $Q$, whose elements are the probabilities to go
to configuration $j$ from configuration $i$, writes
\begin{equation}
Q =
\left(\begin{array}{c|c}
\widetilde{Q}       & \begin{array}{c} \doping^2 \\ \doping \end{array} \\ \hline
\begin{array}{cc} 0 &                   0 \end{array}  &  1 \\
\end{array}\right)
\;,
\end{equation}
where
\begin{equation}
\widetilde{Q}=
\left(\begin{array}{cc}
(1-\doping)^2 & 2\doping(1-\doping) \\
(1-\doping)^2 &  \doping(1-\doping) \\
\end{array}\right)\;.
\end{equation}
The transition probabilities from configuration $3$ are not needed and
we can take this configuration as a trap state. We want to calculate
the distribution of distances to reach configuration $3$
\begin{equation}
\label{eq:proba_distance_cut_def}
\rho(\ell)=\mathbb P\left(X_{\ell+1}=3,X_\ell\neq3,\dots,X_0\neq3\right)\;,
\end{equation}
starting from an initial distribution for $X_0$
\begin{equation}
P_0=\begin{pmatrix}p & 1-p & 0 \end{pmatrix}\;.
\end{equation}
Equation \eqref{eq:proba_distance_cut_def} can be expanded as
\begin{equation}
\label{eq:proba_distance_cut}
\rho(\ell)=\zeta_{\ell}\prod_{n=0}^{\ell-1} (1-\zeta_n)\;,
\end{equation}
where
\begin{equation}
\begin{aligned}
\zeta_{n}
&=\mathbb P\left(X_{n+1}=3 | X_{n}\neq3\right)\\
&=\frac{\doping^2\mathbb P(X_n=1)+\doping\mathbb P(X_n=2)}{\mathbb P(X_n=1)+\mathbb P(X_n=2)}\;.
\end{aligned}
\end{equation}
One can easily show by mathematical induction that
\begin{equation}
\mathbb P(X_n=i)=\left(P_0 Q^n\right)_i\;.
\end{equation}
As a result, we have
\begin{equation}
\zeta_{n}=\frac
{\begin{pmatrix}p & 1-p \end{pmatrix} {\widetilde Q}^n \begin{pmatrix}\doping^2 \\ \doping \end{pmatrix}}
{\begin{pmatrix}p & 1-p \end{pmatrix} {\widetilde Q}^n \begin{pmatrix}1 \\ 1 \end{pmatrix}}\;.
\end{equation}
The distribution \eqref{eq:proba_distance_cut} is not exactly a
geometric law but $\zeta_{n}$ converges really quickly to a constant
\begin{equation}
\zeta=\frac{1}{2}\left(1+\doping-(1-\doping)\sqrt{1+4\doping(1-\doping)}\right)\;,
\end{equation}
independent of $p$, that is, independent of the initial distribution
$P_0$. In the limit $\doping\ll1$, one recovers
$\zeta\simeq3\doping^2$.

The mean cluster size thus has the following low-$\doping$ expansion
\begin{equation}
\bar{\ell} = \frac{1}{3\doping^2}\left[1+2\doping-\frac 4 3\doping^2 +4\doping^3 +\ldots\right]\;.
\end{equation}
On the other limit $\doping \rightarrow 1$, $\bar{\ell} \rightarrow
1$, as expected.

\end{document}